\documentclass[sigconf,anonymous=false]{acmart}

\usepackage{multirow}
\usepackage{subfigure}

\AtBeginDocument{%
  \providecommand\BibTeX{{%
    \normalfont B\kern-0.5em{\scshape i\kern-0.25em b}\kern-0.8em\TeX}}}

\setcopyright{acmcopyright}
\copyrightyear{2022}
\acmYear{2022}
\acmDOI{10.1145/1122445.1122456}

\acmConference[WWW '22]{Proceedings of the ACM Web Conference 2022}{April 25--29, 2022}{Virtual Event, Lyon, France}
\acmBooktitle{Proceedings of the ACM Web Conference 2022 (WWW '22), April 25--29, 2022, Virtual Event, Lyon, France}
\acmPrice{15.00}
\acmDOI{10.1145/3485447.3512166}
\acmISBN{978-1-4503-9096-5/22/04}

\newcommand{\myfont}{\fontsize{8.4pt}{\baselineskip}\selectfont}

\newcommand{\nosection}[1]{\vspace{2pt}\noindent\textbf{#1.}}

\newcommand{\modelname}{\textbf{CFAA}}

\begin{document}

\title{Collaborative Filtering with Attribution Alignment for Review-based Non-overlapped Cross Domain Recommendation}


\author{Weiming Liu, Xiaolin Zheng, Mengling Hu, Chaochao Chen}
\authornote{Chaochao Chen is the corresponding author.}
\affiliation{
  College of Computer Science and Technology, Zhejiang University
  \country{China}\\
\{21831010, xlzheng, humengling, zjuccc\}@zju.edu.cn
}





\begin{abstract}
Cross-Domain Recommendation (CDR) has been popularly studied to utilize different domain knowledge to solve the data sparsity and cold-start problem in recommender systems.
In this paper, we focus on the
\textit{Review-based Non-overlapped Recommendation} (\textit{RNCDR}) problem.
The problem is commonly-existed and challenging due to two main aspects, i.e, there are only positive user-item ratings on the target domain and there is no overlapped user across different domains.
Most previous CDR approaches cannot solve the RNCDR problem well, since (1) they cannot effectively combine review with other information (e.g., ID or ratings) to obtain expressive user or item embedding, (2) they cannot reduce the domain discrepancy on users and items. 
To fill this gap, we propose Collaborative Filtering with Attribution Alignment model (\textbf{\modelname}), a cross-domain recommendation framework for the RNCDR problem. 
\modelname~includes two main modules, i.e., \textit{rating prediction module} and \textit{embedding attribution alignment module}.
The former aims to jointly mine review, one-hot ID, and multi-hot historical ratings to generate expressive user and item embeddings.
The later includes vertical attribution alignment and horizontal attribution alignment, tending to reduce the discrepancy based on multiple perspectives. 
Our empirical study on Douban and Amazon datasets demonstrates that \modelname~significantly outperforms the state-of-the-art models under the RNCDR setting.
\end{abstract}





\maketitle

\setlength{\floatsep}{4pt plus 4pt minus 1pt}
\setlength{\textfloatsep}{4pt plus 2pt minus 2pt}
\setlength{\intextsep}{4pt plus 2pt minus 2pt}
\setlength{\dbltextfloatsep}{3pt plus 2pt minus 1pt}
\setlength{\dblfloatsep}{3pt plus 2pt minus 1pt}
\setlength{\abovecaptionskip}{3pt}
\setlength{\belowcaptionskip}{2pt}
\setlength{\abovedisplayskip}{2pt plus 1pt minus 1pt}
\setlength{\belowdisplayskip}{2pt plus 1pt minus 1pt}

\section{Introduction}








With the advent of digital era, more and more users participant in multiple domains (platforms) for different purposes, e.g., buying books on \emph{Amazon} and reading news on \emph{Flipboard} \cite{new1}.
How to comprehensively utilize the cross domain information to improve the performance of recommendation systems has become a hot topic.
Therefore, Cross Domain Recommendation (CDR) becomes more and more attractive for establishing highly accurate recommendation systems \cite{cdrsurvey,cdrbook,tan,new2}.
%
%
%
%
Most existing CDR models assume that both the source and target domains share the same set of users, which makes it easier to transfer useful knowledge across domains based on these overlapped users.
%
Moreover, most existing CDR models assume the existence of sufficient negative feedback, which limits their applications.

In this paper, we focus on a general problem in CDR where the target domain only has a small proportion of positive ratings but no negative ones. 
Meanwhile users in the source and target domains are totally non-overlapped.
Besides user-item rating information, we also assume the existence of user-item reviews which are commonly used as side-information for alleviating the data sparsity problem \cite{recdan,tdar}.
Specifically, we term this problem as \textit{Review-based Non-overlapped Cross Domain Recommendation} (\textit{RNCDR}), where we aim to transfer knowledge from a relative dense source domain to a sparse target domain to enhance the prediction performance, as shown in Fig.~1. 
The RNCDR problem is widely existed since the target domain is always suffering from the data sparsity problem.
Therefore, the target domain only has a small proportion of observed positive samples without negative ones.
We summarize the main challenges as follows. %
(1) Users/items always have diverse characteristics and preferences which should be well-exploited across domains.
(2) Different domains tend to have varied behavioral patterns \cite{dbwww,ssb}, causing the existence of latent embedding attribution bias and discrepancy \cite{db}. 

%

%
%
%
%

%
Although there have been previous studies on the RNCDR problem \cite{tdar}, they cannot solve it well. 
The current state-of-the-art model on RNCDR is TDAR \cite{tdar} which adopts deep Domain Adversarial Neural Network (DANN) \cite{dann} to align the source and target user/item features.
On the one hand, it fails to integrate the review information with one-hot ID and multi-hot historical rating \cite{pan2008,delf}, leading to less expressive user and item embeddings \cite{conet,atlrec}. 
On the other hand, a deep adversarial network with domain discriminator is unstable and hard to train in practice \cite{dirt-t}.
Therefore, it cannot solve both challenges and leads to poor model performance.
%
%
These two negative aspects make this approach difficult to achieve the RNCDR task. 

\begin{figure}
\centering
\includegraphics[width=\linewidth]{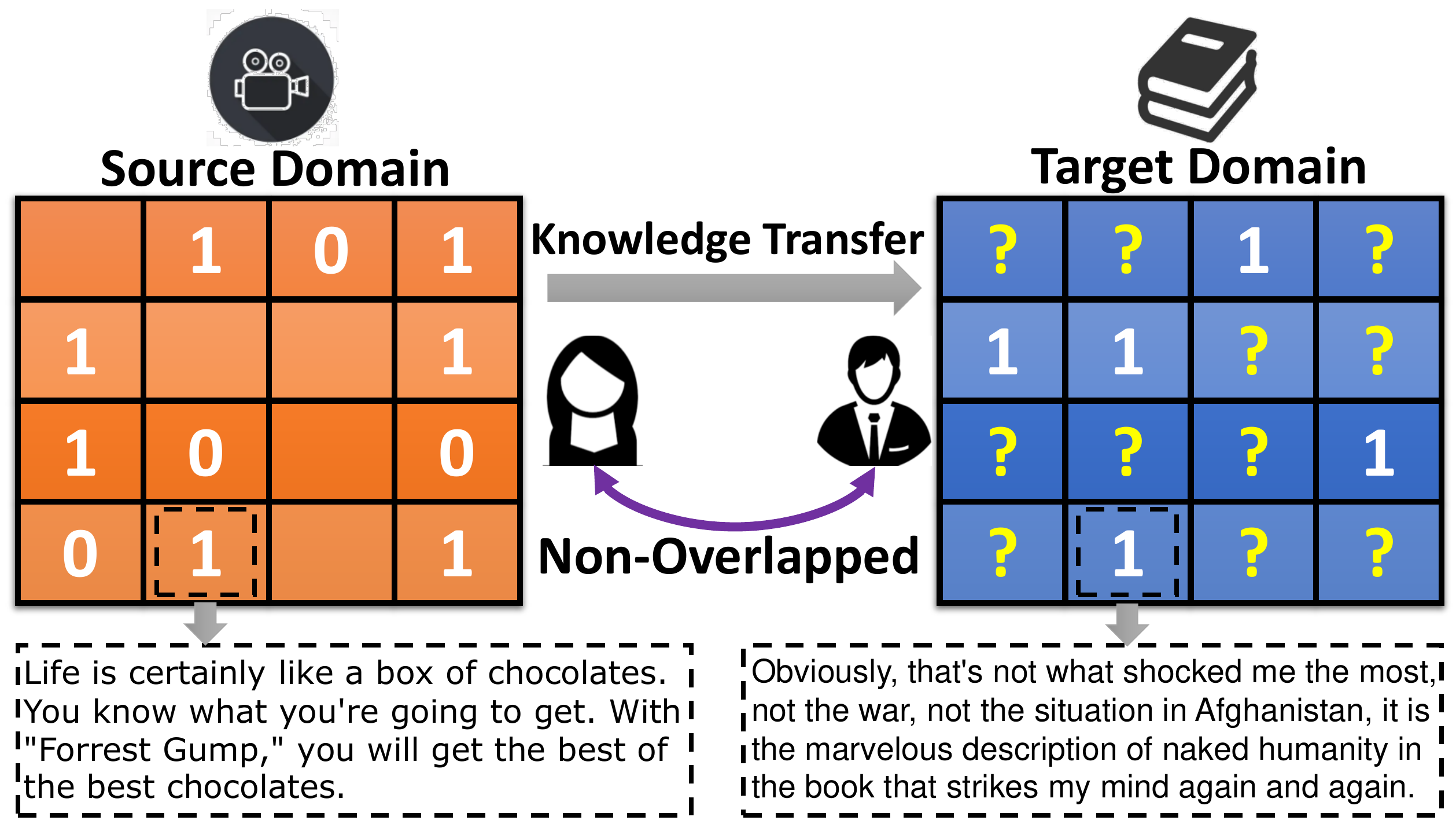}
\caption{The problem of Review-based Non-overlapped Cross Domain Recommendation (RNCDR). }
\label{fig:vade}
\end{figure}

%
%

To address the aforementioned issues, in this paper, we propose \modelname, a cross-domain recommendation framework for the RNCDR problem.
In order to better model user/item embeddings and align the latent embedding attributions across domains for high-quality rating predictions, we utilize two modules in \modelname, i.e., \textit{rating prediction module} and \textit{embedding attribution alignment module}. 
The rating prediction module aims to tackle the first challenge, capturing useful information and generating expressive user and item embeddings.
For this purpose, we fuse the one-hot ID, multi-hot historical ratings, and text-based review information to better capture user and item collaborative preferences.
The embedding attribution alignment module tends to tackle the second challenge, properly reducing the attribution discrepancy and transferring consistent useful knowledge across the source and target domains. 
Specifically, we propose dual perspectives on reducing the discrepancy, i.e., \textit{vertical  attribution alignment} with typical-sample optimal transport and \textit{horizontal attribution alignment} with the adaptation on attribution graph.
Vertical attribution alignment confines the probability distribution among the source and target domains for the corresponding attribution and horizontal attribution alignment tends to exploit the consistent relevant relationship between different attributes across domains. 
These two methods cooperate with each other to complete the alignment task and reduce the domain discrepancy.

We summarize our main contributions as follows:
(1) We propose a novel framework, i.e., \modelname, for the RNCDR problem, which contains rating prediction module and embedding attribution alignment module.
(2) The proposed rating prediction module can efficiently combine the text-based review information with one-hot ID and multi-hot historical rating embedding, and the embedding attribution alignment module equipped with vertical and horizontal alignment can better reduce the domain discrepancy.
(3) Extensive empirical studies on Douban and Amazon datasets demonstrate that \modelname~significantly improves the state-of-the-art models under the RNCDR setting.

\section{Related Work}

\nosection{Traditional Cross Domain Recommendation}
Traditional Cross Domain Recommendation (CDR) emerges as a technique to alleviate the long-standing data sparsity problem in recommendation by assuming the same or partial proportion of overlapped user/item set across domains \cite{cbt,tr2013}. 
Man et al. \cite{emcdr} first proposed to learn the mapping function on the overlapped users across domains.
Zhao et al. \cite{catn} extended the research on the attribution aspects transfer.
Some recent method \cite{dml} even exploited bidirectional latent relations between users and items to obtain more reliable knowledge.
Besides, Hu et al. \cite{conet} further proposed multi-task learning strategy which stitches both source and target information together. 
It is noticeable that the majority of traditional CDR frameworks always assume the existence of overlapped users as the bridge for knowledge sharing across different domains \cite{cdrsurvey}.
However, the source and target domains may not share overlapped users in the real-world applications which limits their application scenario.

%

\nosection{Review-based Non-overlapped Cross Domain Recommendation}
There are some CDR models that can solve the situations where users/items are non-overlapped.
Yang et al. \cite{cbt} was the first to propose the widely used codebook transfer approach for non-overlapped CDR.
However, it cannot be extended to implicit feedback, which strongly limits its generalization capability.
Furthermore, without overlapped users or items act as bridges for knowledge transfer, the recommendation accuracy is still not satisfying \cite{myth}.
Recently, some researchers \cite{deepconn} have propose to utilize review text of users and items to enhance the model performance.
This inspires us to take advantage of these review information as the transferring bridge across different domains.
For example, Wang et al. \cite{recdan} used text-enhanced information for solving both overlapped and non-overlapped CDR problems with deep adversarial learning strategy. 
Yu et al. \cite{tdar} considered a more general situation where the target domain only has positive ratings, and proposed dual adversarial alignment on both user and item embeddings.
However, these models cannot combine the one-hot ID with multi-hot historical rating into the review information, leading to limited model performance. 
%
Moreover, some studies \cite{db} have shown the existence of embedding bias and discrepancy on CDR problem since two domains may always have different expressions.
This enlightens us to transfer useful review knowledge by reducing the embedding bias for better results.

\begin{figure}
\centering
\includegraphics[width=\linewidth]{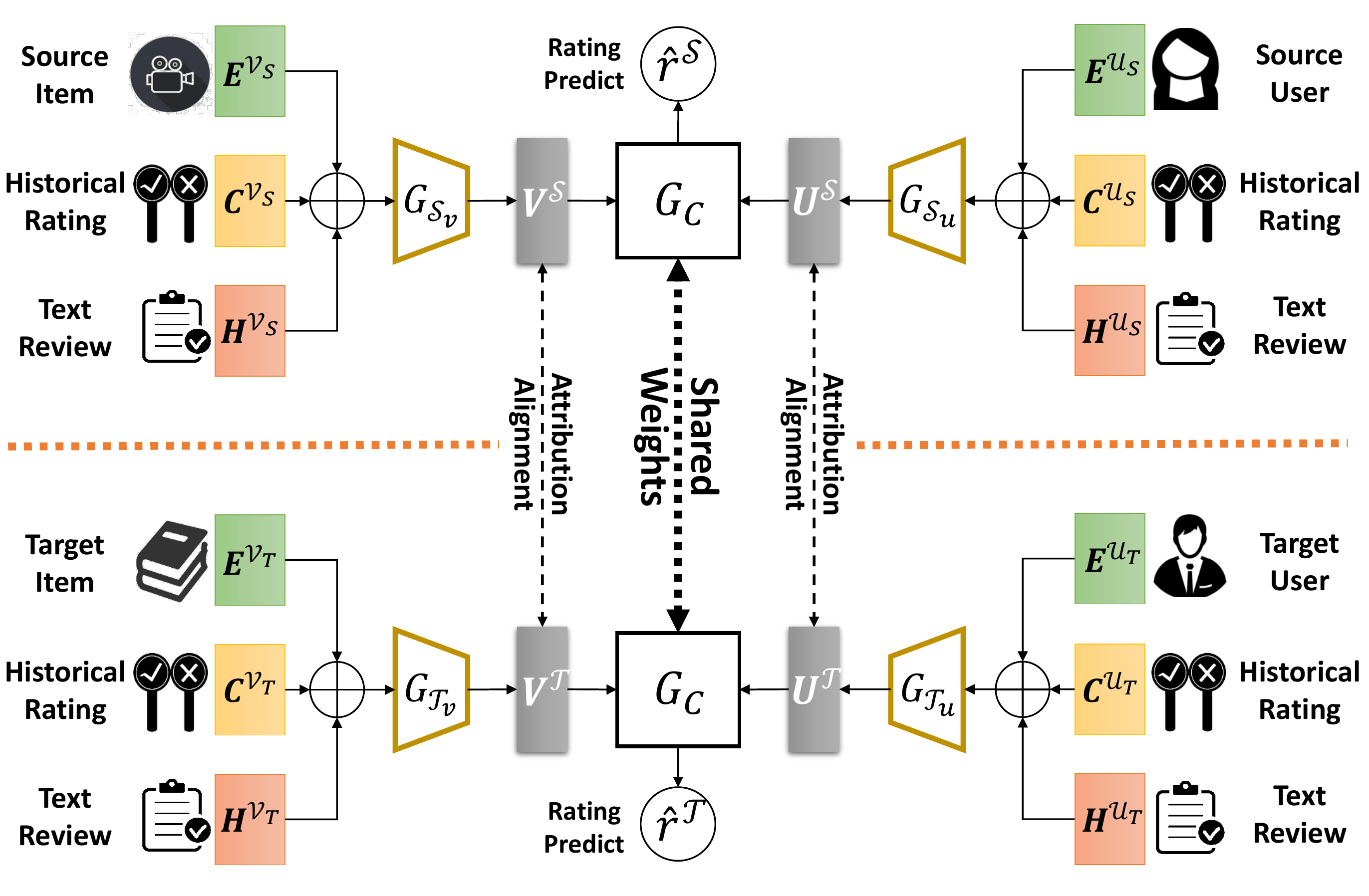}
\vspace{-0.4cm} 
\caption{The basic framework of \modelname.}
\vspace{-0.4cm} 
\label{fig:model}
\end{figure}

\nosection{Domain Adaptation}
Domain adaptation technique was proposed to transfer knowledge from a well-labeled source domain to a target domain without or with less labels \cite{transfer_survey}. 
The most classic method is Maximum Mean Discrepancy (MMD) \cite{tca,mmd} which is based on the mean statistic of samples.
Moreover, some scholars proposed Correlation Alignment (CORAL) \cite{coral,deepcoral} to align the second-order covariance statistics.
More recently, ESAM \cite{esam} extended CORAL with attribution correlation congruence for solving the long-tailed item recommendation problem.
However, performing statistic on mean or covariance may hard to capture the complex and high dimension data characteristics. 
For this, Ganin et al. \cite{dann} proposed Domain Adversarial Neural Network (DANN) which integrated a domain discriminator with adversarial training to align the embeddings across domains. 
Notably that previous review-based CDR models, e.g., TDAR \cite{tdar} and Rec-GAN \cite{recdan}, also adopted adversarial training for aligning the source and target domains.
However, latest researches \cite{dirt-t} pointed out that the origin adversarial training may be unstable under some circumstances which will hurdle the performance. 
In this paper, we propose to enhance statistic-based embedding attribution alignment method by attribution distribution and attribution relevance alignment to exploit more expressive information on complex feature space.

\section{Modeling for \modelname}

First, we describe notations. 
We assume there are two domains, i.e., a source domain $\mathcal{S}$ and a target domain $\mathcal{T}$.
There are $N_{U_S}$ and $N_{U_T}$ users in source and target domains respectively.
There are $N_{V_S}$ and $N_{V_T}$ items in source and target domains respectively.
Let $\boldsymbol{R}^\mathcal{S} \in \mathbb{R}^{N_{U_S} \times N_{V_S}}$ and $\boldsymbol{R}^\mathcal{T} \in \mathbb{R}^{N_{U_T} \times N_{V_T}}$ be the observed source and item rating matrices in $\mathcal{S}$ and $\mathcal{T}$ respectively.
Moreover, let $\boldsymbol{\Omega}^{\mathcal{S}_u}$ and $\boldsymbol{\Omega}^{\mathcal{T}_u}$ be the set of user reviews.
Similarly, let $\boldsymbol{\Omega}^{\mathcal{S}_v}$ and $\boldsymbol{\Omega}^{\mathcal{T}_v}$ be the set of item reviews.
Unlike the traditional cross domain recommendation which has the assumption that both source and target domains share the same set of users, in RNCDR, the source and target users are totally non-overlapped.

Then, we introduce the overview of our proposed \modelname~framework, as is illustrated in Fig.~ \ref{fig:model}.
\modelname~model mainly has two modules, i.e., \textit{rating prediction module} and \textit{embedding attribution alignment module}.
The rating prediction module aims to combine the review information with one-hot ID and multi-hot historical rating to generate expressive user and item embeddings.
The embedding attribution alignment module is supposed to reduce the embedding attribution discrepancy across domains.
We will introduce these two modules in details later.

\subsection{Rating Prediction Module}

Firstly, we provide the details of the rating prediction module.
For convenience, we use the notations and calculation process in the source domain as an example.
For the $i$-th user and the $j$-th item, we define their corresponding one-hot ID vectors as $\boldsymbol{X}^{\mathcal{U}_{S}}_i$ and $\boldsymbol{X}^{\mathcal{V}_{S}}_j$, respectively.
Let $\boldsymbol{R}^\mathcal{S}_{i*}$ and $\boldsymbol{R}^\mathcal{S}_{*j}$ denote the historical rating for the $i$-th user and the $j$-th item.
For the $i$-th user, we use $\boldsymbol{\Omega}^{\mathcal{S}_u}_i$ to denote the corresponding review which includes $\boldsymbol{\Omega}^{\mathcal{S}_u}_i = (\omega^{\mathcal{S}_u}_{i1},\omega^{\mathcal{S}_u}_{i2},\cdots,\omega^{\mathcal{S}_u}_{ik} )$ sentences.
Likewise, we use $\boldsymbol{\Omega}^{\mathcal{S}_v}_j$ to denote the corresponding review for the $j$-th item which includes $\boldsymbol{\Omega}^{\mathcal{S}_v}_j = (\omega^{\mathcal{S}_v}_{j1},\omega^{\mathcal{S}_v}_{j2},\cdots,\omega^{\mathcal{S}_v}_{jl} )$ sentences.
Notably that we adopt the sentence segmentation component (Sentencizer\footnote{https://spacy.io/api/sentencizer}) to split the origin document into several individual sentences in $\boldsymbol{\Omega}^{\mathcal{S}_u}_i$ and $\boldsymbol{\Omega}^{\mathcal{S}_v}_j$.

We adopt a trainable lookup table to exploit the user and item one-hot ID embedding as ${\rm LookUp}(\boldsymbol{X}^{\mathcal{U}}_i) = \boldsymbol{E}^{\mathcal{U}_\mathcal{S}}_i$ and ${\rm LookUp}(\boldsymbol{X}^{\mathcal{V}}_j) = \boldsymbol{E}^{\mathcal{V}_\mathcal{S}}_j$.
We utilize the fully connected layers $F_{S_u}$ and $F_{S_v}$ to obtain user and item behavior embeddings as $ F_{S_u}({\mathcal{R}}^\mathcal{S}_{i*}) = \boldsymbol{C}^{\mathcal{U}_\mathcal{S}}_i$ and $ F_{S_v}({\mathcal{R}}^\mathcal{S}_{*j}) = \boldsymbol{C}^{\mathcal{V}_\mathcal{S}}_j$, respectively. 
Meanwhile we utilize the pre-trained BERT model's penultimate encoder layer to obtain the contextualized word embeddings for each sentence \cite{miller}. 
Then we average every sentence's word embeddings to generate the sentence embedding.
We then average the sentence embeddings of the $i$-th user or $j$-th item as their corresponding review-based embedding $\boldsymbol{H}^{\mathcal{U}_\mathcal{S}}_i$ and $\boldsymbol{H}^{\mathcal{V}_\mathcal{S}}_j$.
Finally, we utilize fully connected layers $G_{S_u}$ and $G_{S_v}$ to obtain the user and item general embedding as $G_{S_u}(\boldsymbol{E}^{\mathcal{U}_\mathcal{S}} \oplus  \boldsymbol{C}^{\mathcal{U}_\mathcal{S}} \oplus \boldsymbol{H}^{\mathcal{U}_\mathcal{S}}) = \boldsymbol{U}^\mathcal{S} \in \mathbb{R}^{N \times D}$ and $G_{S_v}(\boldsymbol{E}^{\mathcal{V}_\mathcal{S}} \oplus  \boldsymbol{C}^{\mathcal{V}_\mathcal{S}} \oplus \boldsymbol{H}^{\mathcal{V}_\mathcal{S}} ) = \boldsymbol{V}^\mathcal{S} \in \mathbb{R}^{N \times D}$ where $N$ denotes batchsize, $D$ denotes the dimension of the latent embedding, 
and $\oplus$ denotes the concatenation operation.
Likewise, we can also obtain the user and item general embeddings on the target domain as $\boldsymbol{U}^\mathcal{T} \in \mathbb{R}^{N \times D}$ and $\boldsymbol{V}^\mathcal{T} \in \mathbb{R}^{N \times D}$ respectively.
After that, we adopt the fully connected layer $G_C$ to predict user-item ratings as $G_C(\boldsymbol{U}^\mathcal{S}, \boldsymbol{V}^\mathcal{S}) = \hat{\boldsymbol{\mathcal{R}}}^\mathcal{S}$ and $G_C(\boldsymbol{U}^\mathcal{T}, \boldsymbol{V}^\mathcal{T}) = \hat{\boldsymbol{\mathcal{R}}}^\mathcal{T}$, respectively.
We further use cross entropy loss $L_{C}$ to minimize the prediction ratings and ground-truth ones as below:
\begin{equation}\nonumber
\begin{aligned}
\label{equ:pairwise_ranking}
L_{C} = &-\sum_{i=1}^N 
({\boldsymbol{\mathcal{R}}}^\mathcal{S}_i \log \hat{\boldsymbol{\mathcal{R}}}^\mathcal{S}_i + (1-{\boldsymbol{\mathcal{R}}}^\mathcal{S}_i) \log (1-\hat{\boldsymbol{\mathcal{R}}}^\mathcal{S}_i) +
{\boldsymbol{\mathcal{R}}}^\mathcal{T}_i \log \hat{\boldsymbol{\mathcal{R}}}^\mathcal{T}_i).
\end{aligned}
\end{equation}
It is noticeable that we only have positive ratings in the target domain.
Through this basic loss function, the network parameters can be adjusted to the training data efficiently. 

\begin{figure*}
    \centering
    \includegraphics[width=\linewidth]{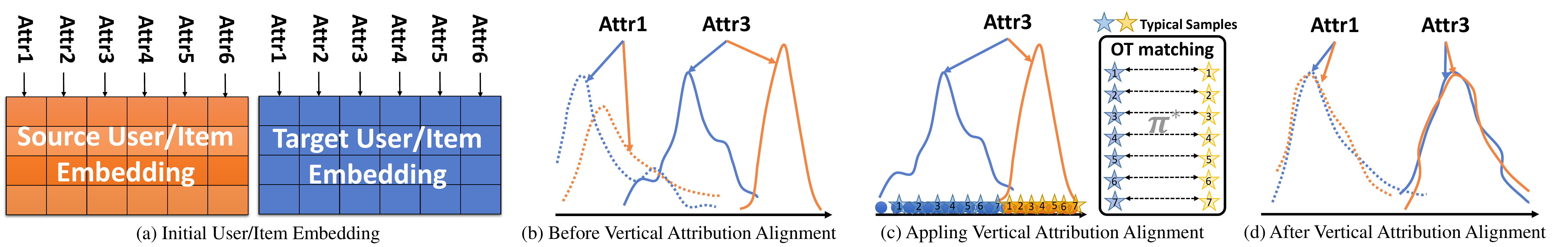}
    \caption{The main procedure of vertical attribution alignment. The orange and blue color denote source and target domains respectively. }
    \label{fig:toy_example1}
\end{figure*}

\subsection{Embedding Attribution Alignment Module}
Although the rating prediction module can provide us a simple and good baseline, it still cannot depict user and item characteristics in the source and target domains well.
The main reason lies in the data sparsity across domains which causes the embedding attribution bias and discrepancy.
Specifically, we denote each dimension of an embedding as a certain kind of \textit{attribution}. 
The attribution discrepancy includes two aspects, i.e., \textit{vertical probability discrepancy} and \textit{horizontal relevance discrepancy} on each attribution, as shown in Fig.~3 and Fig.~4. 
For example, Attr3 in both source and target domains has different \textit{probability distribution} as shown in Fig.~3(b).
Meanwhile, Attr2 and Attr3 have different \textit{relevance strength} across domains, indicating the existence of the relevance discrepancy, as shown in Fig.~4. 
These biases deteriorate the knowledge transfer across domains and may even lead to the negative transfer phenomenon \cite{nt}.
%
%
Therefore, it is essential to reduce the embedding attribution discrepancy for better knowledge transfer.
The embedding attribution discrepancy module consists of two main algorithms, i.e., \textit{Vertical Attribution Alignment} and \textit{Horizontal Attribution Alignment}.
These two algorithms proceed from different perspectives, and they complement each other to solve the embedding attribution discrepancy problem.

\subsubsection{Vertical Attribution Alignment}
We first introduce the attribution distribution alignment. 
It is reasonable to assume that each dimension of user (item) embeddings has certain meaningful information like occupation, income, hobby (style, theme, brand), and etc.
Although we cannot directly decipher what is the exact meaning of each dimension, deep analysis of these latent embedding attributions can still be helpful for recommendation \cite{esam}.
Each latent attribution distribution should be consistent with the source and target domains in order to enhance the model performance.
However, due to the data sparsity among the CDR problem, there always exists the \textit{vertical distribution discrepancy} between the source and target domains, as shown in Fig.~3. 
For example, Attr1 and Attr3 in Fig.~3(b) have different probability distribution across domains with a certain domain gap.
Vertical attribution discrepancy seriously hurdles the knowledge transfer across domains, which not only raises the training difficulty but also leads to the scattered target space with domain discrepancy. 
Vertical attribution alignment mainly has two steps, i.e., \textit{typical sample selection step} and \textit{optimal transport matching step}. 
The former can figure out the proxy data samples and filter the irrelevant noise data, 
while the later can align these typical samples on the each attribution across different domains.

\nosection{Typical Sample Selection}
We first introduce the \textit{typical sample selection algorithm}, which aims to find $K$ typical samples $\boldsymbol{M}^{\mathcal{X}}_{Z} \in \mathbb{R}^{K \times D}$ in both source and target domains where the $q$-th column $(\boldsymbol{M}^{\mathcal{X}}_{Z})_{*q} \in \mathbb{R}^{K \times 1}$ represents the typical samples on the $q$-th attribution. 
Let $\mathcal{X} = \{\mathcal{S},\mathcal{T}\}$ denote the domain index, and $\boldsymbol{Z} = \{\boldsymbol{U},\boldsymbol{V}\}$ denote the user or item set. 
Inspired by \cite{convex,kmm}, we formulate the typical sample selection optimization problem as:
\begin{equation}
\myfont{
\begin{aligned}
\label{equ:multiple_proxies3}
\min \ell_q = \sum_{i=1}^{N} \sum_{j=1}^{K}  (\boldsymbol{\Psi}^{Z^{\mathcal{X}}}_q )_{ij}\left(\boldsymbol{Z}_{iq}^\mathcal{X} - (\boldsymbol{M}^{\mathcal{X}}_Z)_{jq} \right)^2 + \alpha \cdot \boldsymbol{\mathcal{R}}\left((\boldsymbol{\Psi}^{Z^{\mathcal{X}}}_q)_{ij} \right)\\
s.t.\,\, (\boldsymbol{\Psi}^{Z^{\mathcal{X}}}_q )_{i} \boldsymbol{1} = 1, \boldsymbol{\mathcal{R}}\left((\boldsymbol{\Psi}^{Z^{\mathcal{X}}}_q)_{ij} \right) = (\boldsymbol{\Psi}^{Z^{\mathcal{X}}}_q)_{ij} \log (\boldsymbol{\Psi}^{Z^{\mathcal{X}}}_q)_{ij}, (\boldsymbol{\Psi}^{Z^{\mathcal{X}}}_q)_{ij} > 0,
\end{aligned}
}
\end{equation}
where $\boldsymbol{\Psi}^{Z^{\mathcal{X}}}_{q} \in \mathbb{R}^{N \times K}$ be the similarity matrix between the data samples and the typical proxies on the $q$-th attribution.
%
%
%
%
%
The nonnegative entropy norm term $\boldsymbol{\mathcal{R}}((\boldsymbol{\Psi}^{Z^{\mathcal{X}}}_q)_{ij} ) = (\boldsymbol{\Psi}^{Z^{\mathcal{X}}}_q )_{ij} \log (\boldsymbol{\Psi}^{Z^{\mathcal{X}}}_q )_{ij}$ is set to avoid trivial solution with $\alpha$ denoting the regularization strength \cite{convex}.
%
%
We will provide the optimization details on the typical sample selection algorithm in Appendix A.
In short, alternatively updating $\boldsymbol{M}^{\mathcal{X}}_Z$ and $\boldsymbol{\Psi}^{Z^{\mathcal{X}}}_{q}$ can solve Equation (1) efficiently as:
\begin{equation}
\begin{aligned}
\label{equ:find_s1}
(\boldsymbol{\Psi}^{Z^{\mathcal{X}}}_q)_{ij} = \frac{\exp \left( -\zeta_{ijq}/ \alpha \right)}{\sum_{k=1}^K \exp ( -\zeta_{ikq}/ \alpha )},\,(\boldsymbol{M}^{\mathcal{X}}_Z)_{jq} = \frac{\sum_{i=1}^{N} (\boldsymbol{\Psi}^{Z^{\mathcal{X}}}_q )_{ij}\boldsymbol{Z}_{iq}^\mathcal{X} }{\sum_{i=1}^{N} (\boldsymbol{\Psi}^{Z^{\mathcal{X}}}_q)_{ij}},
\end{aligned}
\end{equation}
where $\zeta_{ijq} = (\boldsymbol{Z}_{iq}^\mathcal{X} - (\boldsymbol{M}^{\mathcal{X}}_Z )_{jq} )^2$.
Since this problem is convex, we can obtain the stable solution of $\boldsymbol{\Psi}^{Z^{\mathcal{X}}}_{q}$ and $\boldsymbol{M}^{\mathcal{X}}_Z$ through iterations.

\begin{figure*}
    \centering
    \includegraphics[width=\linewidth]{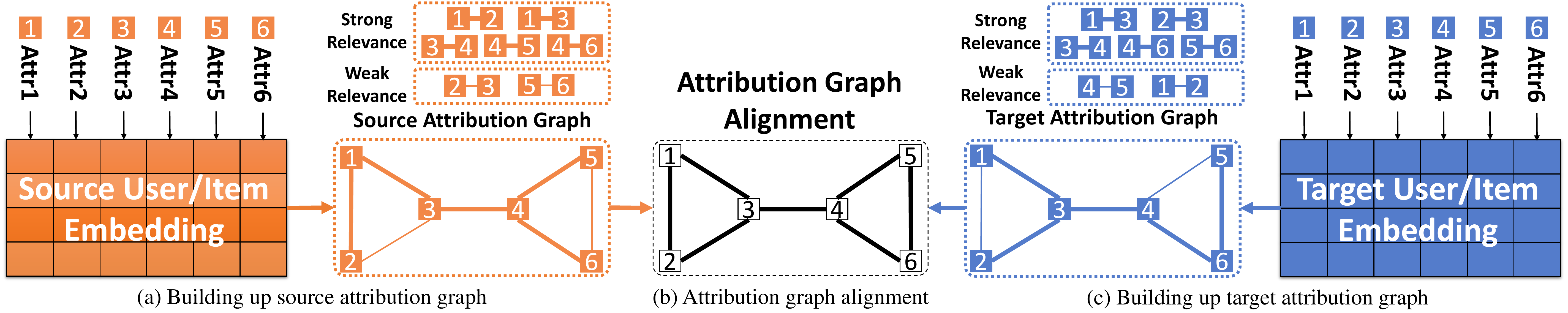}
    \caption{The main procedure of horizontal attribution alignment.}
    \label{fig:toy_example2}
\end{figure*}

\nosection{Optimal Transport Alignment}
In order to better model the attribution distribution between the source and target domains, we adopt the optimal transport technique \cite{deepjdot}.
Optimal transport is based on Kantorovich problem \cite{angenent2003minimizing}, seeking for a general coupling $\boldsymbol{\pi}^{Z}_{q} \in \mathcal{X}(\mathcal{S}_q,\mathcal{T}_q)$ between $\mathcal{S}_q$ and $\mathcal{T}_q$ on the $q$-th attribution:
\begin{equation}\nonumber
\begin{aligned}
\label{loss_ot_def1}
\hat{\boldsymbol{\pi}}^{Z}_{q} = \mathop{\arg\min} \int \mathcal{M}((\boldsymbol{M}^{\mathcal{S}}_Z)_{*q},(\boldsymbol{M}^{\mathcal{T}}_Z )_{*q} ) \,d\boldsymbol{\pi}^{Z}_{q} ((\boldsymbol{M}^{\mathcal{S}}_Z )_{*q},(\boldsymbol{M}^{\mathcal{T}}_Z )_{*q} ).
\end{aligned}
\end{equation}
The cost function matrix $\mathcal{M}((\boldsymbol{M}^{\mathcal{S}}_Z)_{*q},(\boldsymbol{M}^{\mathcal{T}}_Z )_{*q} )$ denotes the cost to move probability mass from  $(\boldsymbol{M}^{\mathcal{S}}_Z)_{*q}$ to $(\boldsymbol{M}^{\mathcal{T}}_Z)_{*q}$ where $(\boldsymbol{M}^{\mathcal{S}}_Z)_{*q}$ and $(\boldsymbol{M}^{\mathcal{T}}_Z)_{*q}$ denote the typical selected samples on the $q$-th attribution for source and target domain respectively. 
The discrete optimal transport formulation can be expressed as:
\begin{equation}
\begin{aligned}
\label{loss_ot_def2}
\hat{\boldsymbol{\pi}}^{Z}_{q} = \mathop{\arg\min} \left[ \langle \boldsymbol{\pi}^{Z}_{q},\mathcal{M}\rangle_{F} + \epsilon \mathcal{R}(\boldsymbol{\pi}^{Z}_{q}) \right],
\end{aligned}
\end{equation}
where $\hat{\boldsymbol{\pi}}^{Z}_{q} \in \mathbb{R} ^ {K \times K}$ is the ideal coupling matrix between the source typical samples $(\boldsymbol{M}^{\mathcal{S}}_Z)_{*q}$ and the target typical samples $(\boldsymbol{M}^{\mathcal{T}}_Z)_{*q}$. 
The matching matrix follows the constraint of $\boldsymbol{1}_K\boldsymbol{\pi}^{Z}_{q} = \boldsymbol{1}_K \left(\boldsymbol{\pi}^{Z}_{q}\right)^T = \frac{1}{K}\boldsymbol{1}_K$.
The second term $\mathcal{R}(\boldsymbol{\pi}^{Z}_{q})$ is the regularization term and $\epsilon$ is a hyper-parameter to balance the entropy regularization and matching loss. 
The matrix $\mathcal{{M}} \in \mathbb{R}^{K \times K}$ denotes the pairwise distance as $\mathcal{M}[i][j] = ||(\boldsymbol{M}^{\mathcal{S}}_Z)_{iq} - (\boldsymbol{M}^{\mathcal{T}}_Z)_{jq}||_2^2$.
Therefore, we can calculate the optimal transport distance $d_O$ as:
\begin{equation}
\begin{aligned}
\label{loss_ot_def3}
d_O((\boldsymbol{M}^{\mathcal{S}}_Z)_{*q},(\boldsymbol{M}^{\mathcal{T}}_Z )_{*q}) = \frac{1}{K^2} \sum_{i=1}^K \sum_{j=1}^K \left[\hat{\boldsymbol{\pi}}^{Z}_{q}\right]_{ij}||(\boldsymbol{M}^{\mathcal{S}}_Z)_{iq} - (\boldsymbol{M}^{\mathcal{T}}_Z)_{jq}||_2^2. 
\end{aligned}
\end{equation}
In summary, we propose the attribution distribution distance alignment loss as below:
\begin{equation}
\begin{aligned}
\label{loss_ot_def4}
L_O = \frac{1}{D} \sum_{q=1}^D \left[ d_O((\boldsymbol{M}^{\mathcal{S}}_U)_{*q},(\boldsymbol{M}^{\mathcal{T}}_U )_{*q}) + d_O((\boldsymbol{M}^{\mathcal{S}}_V)_{*q},(\boldsymbol{M}^{\mathcal{T}}_V )_{*q}) \right].
\end{aligned}
\end{equation}
%
%
Take Fig.~3 for example, we first collect the typical samples of Attr3, e.g., the stars marked with numbers in Fig.~3(c), in both source and target domains.
Then we adopt the Optimal Transport (OT) to align these typical samples across domains.
Afterward, we minimize the distance between the matched typical samples.
Finally, we can align the attribution probability distribution in Fig.~3(d).

\subsubsection{Horizontal Attribution Alignment}
Then we introduce the horizontal attribution alignment algorithm.
Previous researches have pointed out that aligning the corresponding relevant attribution relationship, e.g., adopting the correlation alignment with covariance matrix in ESAM \cite{esam}, can better enhance the model performance.
However, covariance is hard to capture the complex and nonlinear hidden relationships between different attributions under the RNCDR problem \cite{jdda}. 
Therefore, we propose horizontal attribution alignment with attribution subspace modelling and attribution graph alignment  methods.
We first use attribution subspace modelling to build the attribution graph, exploiting the hierarchical and topological structure between different attributions. 
Then we propose attribution graph alignment to align attribution graphs in source and target domains with Wasserstein distance metric. 

\nosection{Attribution Subspace Modelling}
SLIM (Sparse LInear Methods) \cite{slim,lorslim} have been widely adopted in recommendation systems due to its good performance.
It computes the item-item relations with statistical learning using the corresponding coefficient matrix.
Meanwhile, one can even adopt it to measure the embedding attribution-attribution relations according to the following optimization problem:
\begin{equation}
\begin{aligned}
\label{equ:slim_1x}
\min_{\boldsymbol{B}^\mathcal{X}_Z} \frac{1}{2}\left| \left|\boldsymbol{Z}^\mathcal{X} - \boldsymbol{Z}^\mathcal{X}\boldsymbol{B}^\mathcal{X}_Z \right|\right|_2^2 + \nu \left|\left|\boldsymbol{B}^\mathcal{X}_Z \right|\right|_*, s.t.\,\, {\rm diag}\left(\boldsymbol{B}^\mathcal{X}_Z \right) = 0,
\end{aligned}
\end{equation}
where $\mathcal{X} = \{\mathcal{S},\mathcal{T}\}$ denotes the domain index, $\boldsymbol{Z} = \{\boldsymbol{U},\boldsymbol{V}\}$ denotes the user and item set, 
${\rm diag}(\boldsymbol{B}^\mathcal{X}_Z ) = 0$ is a constant that avoids the trivial solution,
$\nu$ is the balance hyper parameter, and $||\cdot||_*$ is the nuclear-norm which can make the matrix to become low rank \cite{lorslim}. 
The low rank constraint can also enhance the robustness and generalization of $\boldsymbol{B}^\mathcal{X}_Z$.
Considering that $||\boldsymbol{B}^\mathcal{X}_Z ||_* = {\rm Tr}\left(\sqrt{(\boldsymbol{B}^\mathcal{X}_Z)^T \boldsymbol{B}^\mathcal{X}_Z} \right)$, the original optimization problem Equation \eqref{equ:slim_1x} can be rewritten as: 
\begin{equation}
\begin{aligned}
\label{equ:slim_2x}
\min_{\boldsymbol{B}^\mathcal{X}_Z,\boldsymbol{\Phi}} \frac{1}{2}\left| \left|\boldsymbol{Z}^\mathcal{X} - \boldsymbol{Z}^\mathcal{X}\boldsymbol{B}^\mathcal{X}_Z \right|\right|_2^2 + \nu {\rm Tr}\left( (\boldsymbol{B}^\mathcal{X}_Z)^T \boldsymbol{\Phi} \boldsymbol{B}^\mathcal{X}_Z  \right)\\
s.t.\,\, {\rm diag}\left(\boldsymbol{B}^\mathcal{X}_Z\right) = 0, \boldsymbol{\Phi} = \left( \boldsymbol{B}^\mathcal{X}_Z (\boldsymbol{B}^\mathcal{X}_Z)^T \right)^{-\frac{1}{2}}.
\end{aligned}
\end{equation}
Alternatively updating $\boldsymbol{B}^\mathcal{X}_Z$ and $\boldsymbol{\Phi}$ can solve Equation \eqref{equ:slim_2x} efficiently. 
For $\boldsymbol{B}^\mathcal{X}_Z$, it has the following closed-form solution:
\begin{equation}
\begin{aligned}
\left(\boldsymbol{B}^\mathcal{X}_Z\right)_{ij} = 
\left\{
\begin{aligned}
&0, \quad \quad \quad \quad i = j\\
& -\frac{\boldsymbol{\Theta}_{ij}}{\boldsymbol{\Theta}_{jj}},\quad \rm{Others}.
\end{aligned}
\right.
\end{aligned}
\end{equation}
where $\boldsymbol{\Theta} = ( ( \boldsymbol{Z}^\mathcal{X})^T \boldsymbol{Z}^\mathcal{X} +  \nu (\boldsymbol{\Phi} + \boldsymbol{\Phi}^T) )^{-1}$.
After we have updated $\boldsymbol{B}^\mathcal{X}_Z$, we fix it as a constant and update $\boldsymbol{\Phi}$ through the equality constraint $\boldsymbol{\Phi} = ( \boldsymbol{B}^\mathcal{X}_Z (\boldsymbol{B}^\mathcal{X}_Z)^T)^{-\frac{1}{2}}$.
By utilizing the iterative updating method until it converges, we can obtain the results of $\boldsymbol{B}^\mathcal{X}_Z$ and $\boldsymbol{\Phi}$.
%
%
The optimization details will be provided in Appendix B.
Notably that $\boldsymbol{B}^\mathcal{X}_Z$ may be asymmetric and contain negative values, we build up the attribution graph by taking the average value of $|\boldsymbol{B}^\mathcal{X}_Z|$ and $|(\boldsymbol{B}^\mathcal{X}_Z)^T|$ which can be depicted as $\boldsymbol{A}^\mathcal{X}_Z = (|\boldsymbol{B}^\mathcal{X}_Z| + |(\boldsymbol{B}^\mathcal{X}_Z)^T)/2$, 
%
%
where $(\boldsymbol{A}^\mathcal{X}_Z)_{ij}$ depicts the attribution similarity on the $i$-th and $j$-th attribution \cite{rosc}.
Meanwhile $\boldsymbol{A}^\mathcal{X}_Z$ can be viewed as the graph adjacent matrix where each node denotes the corresponding attribution.
Therefore, we can establish the attribution graph to represent the topology structure among these attributions, as shown in Fig.~4(a) and Fig.~4(c) in the source and target domains.
%

\begin{figure}
    \centering
    \includegraphics[width=\linewidth]{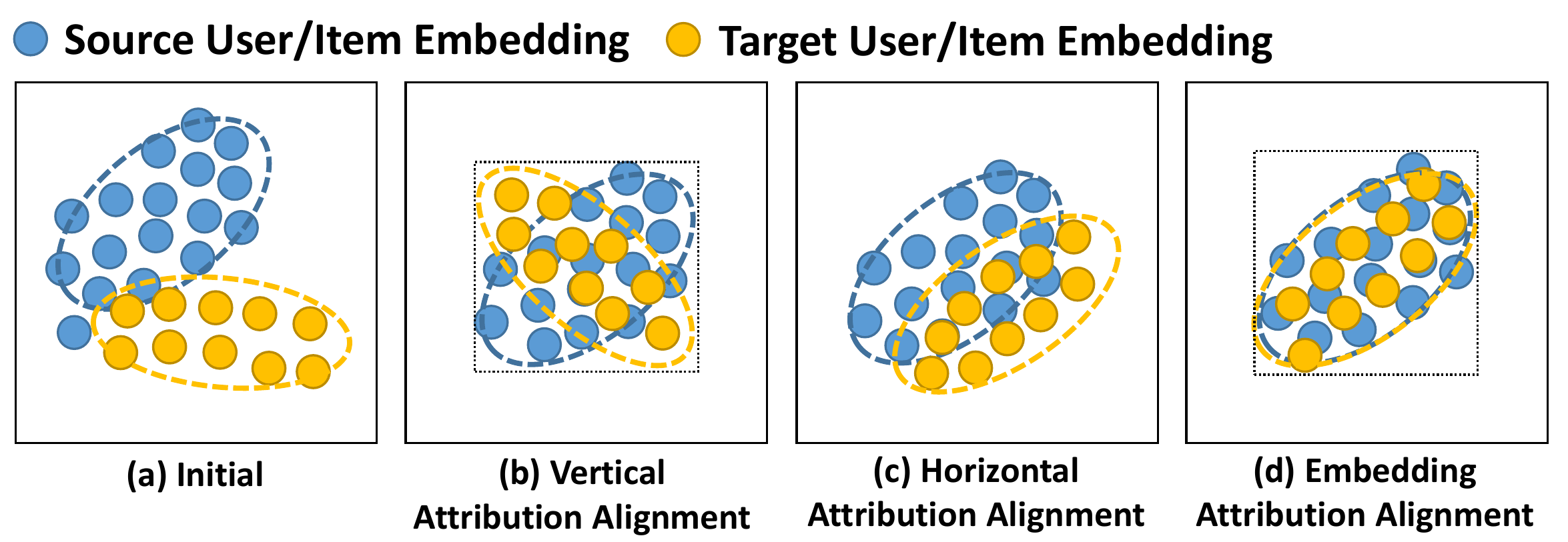}
    \caption{The illustrations of attribution alignment. }
    \label{fig:toy_example3}
\end{figure}

\nosection{Attribution Graph Alignment}
After we obtain the attribution graph through the adjacent matrix $\boldsymbol{A}^\mathcal{X}_Z$, we can match the source and target attribution graph across domains in Fig.~4(b).
To start with, we first calculate the corresponding Laplacian matrix $\boldsymbol{L}^\mathcal{S}_Z = \boldsymbol{D}^\mathcal{S}_Z - \boldsymbol{A}^\mathcal{S}_Z$ and 
$\boldsymbol{L}^\mathcal{T}_Z = \boldsymbol{D}^\mathcal{S}_Z - \boldsymbol{A}^\mathcal{S}_Z$ through the origin adjacent matrix where $\boldsymbol{D}^\mathcal{S}_Z$ and $\boldsymbol{D}^\mathcal{T}_Z$ denote the degree matrix on graph.

According to previous research \cite{got2}, a graph can be interpreted as a multivariate normal distribution. 
Specifically, the mean of the multivariate normal distribution is zero and the covariance is the inverse Laplacian matrix $(\boldsymbol{L}^\mathcal{S}_Z)^+$ and $(\boldsymbol{L}^\mathcal{T}_Z)^+$, shown as below:
\begin{equation}
\begin{aligned}
\mathbb{P}(\boldsymbol{\mathcal{G}}^\mathcal{S}_Z) = \mathcal{N}(0,(\boldsymbol{L}^\mathcal{S}_Z)^+),\quad \mathbb{P}(\boldsymbol{\mathcal{G}}^\mathcal{T}_Z) = \mathcal{N}(0,(\boldsymbol{L}^\mathcal{T}_Z)^+).
\end{aligned}
\end{equation}
%
%
The above formulation has been further used in many graph inference algorithms to represent the graph characteristics.
Therefore, we can calculate the attribution graph distance and reduce the discrepancy based on the formulation. 
We adopt the Wasserstein distance \cite{ws} to measure the distance between different Gaussians distributions, which is shown as: 
\begin{equation}
\begin{aligned}
\label{equ:was}
&d_W( \mathcal{N}(0,(\boldsymbol{L}^\mathcal{S}_Z)^+), \mathcal{N}(0,(\boldsymbol{L}^\mathcal{T}_Z)^+))\\&= {\rm Tr}\left((\boldsymbol{L}^\mathcal{S}_Z)^+ + (\boldsymbol{L}^\mathcal{T}_Z)^+ - 2((\boldsymbol{L}^\mathcal{S}_Z)^+)^{\frac{1}{2}} (\boldsymbol{L}^\mathcal{T}_Z)^+ ((\boldsymbol{L}^\mathcal{S}_Z)^+)^{\frac{1}{2}} )^{\frac{1}{2}} \right).
\end{aligned}
\end{equation}
Meanwhile the Wasserstein distance on graph is more sensitive than traditional Frobenius norm distance which can detect more subtle modification \cite{got}. 
%
%
%
%
%
Therefore we propose the attribution map distance alignment loss as below:
\begin{equation}
\begin{aligned}
L_{A} = d_W\left( \mathbb{P}(\boldsymbol{\mathcal{G}}^\mathcal{S}_U) , \mathbb{P}(\boldsymbol{\mathcal{G}}^\mathcal{T}_U) \right) + d_W \left( \mathbb{P}(\boldsymbol{\mathcal{G}}^\mathcal{S}_V) , \mathbb{P}(\boldsymbol{\mathcal{G}}^\mathcal{T}_V) \right).
\end{aligned}
\end{equation}
The relevance between different attributions will finally meet the consensus through attribution graph alignment, as shown in Fig.~4(b).
In summary, vertical and horizontal attribution alignment are both indispensable, as shown in Fig.~5.
With the vertical attribution distribution alignment, one can just align the marginal distribution across domains, as shown in Fig.~5(b). 
Otherwise, the horizontal attribution alignment can only align the attribution relationship but cannot reduce the distribution discrepancy as shown in Fig.~5(c). 
Therefore, two methods can complement each other and work together to complete the task, as shown in Fig.~5(d).

\subsection{Putting Together}
The total loss of \modelname~could be obtained by combining the losses of the rating prediction module and the embedding attribution alignment module.
That is, the loss of \modelname~is given as:
\begin{equation}
\begin{aligned}
\label{equ:total}
L_{\modelname} = L_{C} + \lambda_{O} L_{O} + \lambda_{A} L_{A}, 
\end{aligned}
\end{equation}
where $\lambda_{O}$ and $\lambda_{A}$ are hyper-parameters to balance different type of losses.
By doing this, \modelname~can not only model the source user-item interactions, but also reduce the embedding bias and discrepancy on user and item across domains.

%
%



\section{Empirical Study}

\begin{table*}[!t]
\centering
\caption{Experimental results on Douban and Amazon datasets. }
\label{tab:douban2}
\begin{tabular}{ccccccccccccc}
\toprule
\multirow{2}{*}{}
& \multicolumn{3}{c}{(Amazon) Movie$\rightarrow$Video}
& \multicolumn{3}{c}{(Amazon) Movie$\rightarrow$Music}
& \multicolumn{3}{c}{(Amazon) Movie$\rightarrow$Clothes}
& \multicolumn{3}{c}{(Amazon) Book$\rightarrow$Video}\\
\cmidrule(lr){2-4} \cmidrule(lr){5-7} \cmidrule(lr){8-10} \cmidrule(lr){11-13}
&HR &Recall &NDCG
&HR &Recall &NDCG
&HR &Recall &NDCG
&HR &Recall &NDCG\\

\midrule
DeepCoNN & .1459 & .1338 & .0944 & .1476 & .1588 & .0625 & .1039 & .1135 & .0517   & .1274 & .1816 & .0798\\ 
NARRE & .1538 & .1413 & .1032 & .1573 & .1707 & .0948 & .1346 & .1270 & .0654   & .1491 & .1902 & .0870\\
Rec-GAN & .1601 & .1692 & .1065 & .1787 & .1876 & .1193 & .1560 & .1402 & .0726   & .1597 & .1969 & .1033\\
TDAR & .1782 & .1786 & .1197 & .1915 & .2043 & .1485 & .1674 & .1461 & .0802   & .1709 & .2050 & .1167\\
ESCOFILT & .1896 & .1977 & .1304 & .2024 & .2115 & .1567 & .1751 & .1504 & .0893   & .1835 & .2198 & .1281\\
DARec & .2044 & .2220 & .1651 & .2462 & .2417 & .1709 & .1846 & .1708 & .0995  & .2096 & .2427 & .1589\\
ESAM & .2067 & .2275 & .1696 & .2408 & .2391 & .1680 & .1870 & .1715 & .1012   & .2113 & .2402 & .1561\\
\midrule 
\modelname-Base & .2013 & .2168 & .1569 & .2316 & .2301 & .1624 & .1832 & .1686 & .0965   & .2048 & .2349 & .1522\\
\modelname-V & .2083 & .2263 & .1683 & .2440 & .2514 & .1736 & .1865 & .1713 & .1004   & .2133 & .2486 & .1679\\
\modelname-H & .2105 & .2205 & .1722 & .2391 & .2468 & .1705 & .1891 & .1735 & .1028   & .2167 & .2513 & .1705\\
\modelname & \textbf{.2190} & \textbf{.2414} & \textbf{.1808} & \textbf{.2552} & \textbf{.2637} & \textbf{.1901} & \textbf{.1944} & \textbf{.1798} & \textbf{.1110}   & \textbf{.2250} & \textbf{.2605} & \textbf{.1816}\\
\midrule

\multirow{2}{*}{}
& \multicolumn{3}{c}{(Amazon) Book$\rightarrow$Music}
& \multicolumn{3}{c}{(Amazon) Book$\rightarrow$Clothes}
& \multicolumn{3}{c}{(Douban) Movie$\rightarrow$Music}
& \multicolumn{3}{c}{(Douban) Book$\rightarrow$Music}
\\
\cmidrule(lr){2-4} \cmidrule(lr){5-7} \cmidrule(lr){8-10} \cmidrule(lr){11-13}
&HR &Recall &NDCG
&HR &Recall &NDCG
&HR &Recall &NDCG
&HR &Recall &NDCG\\

\midrule
DeepCoNN & .1169 & .1624 & .0706 & .0939 & .1085 & .0439 & .1868 & .1540 & .0787   & .1485 & .1363 & .0571\\ 
NARRE & .1382 & .1696 & .0933 & .1148 & .1137 & .0504 & .2037 & .1669 & .0944   & .1576 & .1580 & .0962\\
Rec-GAN & .1599 & .1817 & .1228 & .1261 & .1206 & .0613 & .2204 & .1782 & .1193   & .1798 & .1636 & .0984\\
TDAR & .1708 & .2032 & .1414 & .1465 & .1291 & .0688 & .2365 & .2016 & .1328   & .1913 & .1792 & .1109\\
ESCOFILT & .1784 & .2108 & .1570 & .1587 & .1369 & .0745 & .2396 & .2103 & .1399   & .2057 & .1904 & .1250\\
DARec & .1897 & .2295 & .1840 & .1694 & .1501 & .0873 & .2562 & .2415 & .1638   & .2203 & .2101 & .1405\\
ESAM & .1915 & .2321 & .1909 & .1712 & .1530 & .0897 & .2589 & .2366 & .1544   & .2219 & .2133 & .1393\\
\midrule 
\modelname-Base & .1860 & .2265 & .1761 & .1670 & .1488 & .0854 & .2530 & .2327 & .1502   & .2194 & .2075 & .1368\\
\modelname-V & .1924 & .2340 & .1912 & .1699 & .1522 & .0910 & .2628 & .2410 & .1713   & .2250 & .2146 & .1421\\
\modelname-H & .1913 & .2319 & .1876 & .1725 & .1543 & .0941 & .2605 & .2424 & .1690   & .2236 & .2137 & .1434\\
\modelname & \textbf{.2021} & \textbf{.2493} & \textbf{.2015} & \textbf{.1782} & \textbf{.1595} & \textbf{.1016} & \textbf{.2715} & \textbf{.2548} & \textbf{.1821}   & \textbf{.2321} & \textbf{.2210} & \textbf{.1496}\\
\bottomrule

\end{tabular}
\end{table*}

In this section, we conduct experiments on several real-world datasets to answer the following questions: 
(1) \textbf{RQ1}: How does our approach perform compared with the state-of-the-art CDR methods?
(2) \textbf{RQ2}: How do the vertical and horizontal attribution alignments contribute to performance improvement? 
(3) \textbf{RQ3}: How does the performance of \modelname~vary with different values of the hyper-parameters?

\subsection{Datasets and Tasks} 
We conduct extensive experiments on two popularly used real-world datasets, i.e., \textit{Douban} and \textit{Amazon}.
First, the \textbf{Douban} dataset \cite{dtcdr,gadtcdr} has three domains, i.e., Book, Music, and Movie. 
Second, the \textbf{Amazon} dataset \cite{catn,amazon} has five domains, i.e., Movies and TV (Movie), Books (Book), CDs and Vinyl (Music), Instant Videos (Video) and Clothes (Clothes).
Both datasets have user-item ratings and reviews. 
The detailed statistics of these datasets after pre-process are shown in Appendix C.
We select the relative large datasets (e.g., \textbf{Amazon Movie}, \textbf{Douban Book}) as the source domains and the rest as the target domains.
We remove the users and items less than 30 records to increase the density following existing research in the source domain \cite{tdar}. 
Meanwhile we also delete some part of interactions in the target datasets to make them more sparse.
For each datasets, we binarize the ratings to 0 and 1.
Specifically, we take the ratings higher or equal to 4 as 1 and others as 0. 
Therefore we conduct several tasks that transferring the useful knowledge from the source to the target domains.
Notably that it includes both easy and hard tasks, e.g., the transfer task between \textbf{Amazon Movie} $\rightarrow$ \textbf{Amazon Video} is easy since they are rather similar, while \textbf{Amazon Movie} $\rightarrow$ \textbf{Amazon Clothes} is hard because they are different.

\subsection{Experiment Settings}

We randomly divide the observed source and target data into training, validation, and test sets with a ratio of 8:1:1.
Users and items are selected to be both non-overlapped across domains. 
We set batch size $N=256$ for both the source and target domains.
The latent embedding dimension is set to $D = 300$.
For the English reviews on Amazon, we apply the pretrained transformer based on RoBERTa-Large for semantic textual similarity task \cite{roberta}.
For the Chinese reviews on Douban, we apply the pretrained Chinese BERT \cite{cbert}.
We set $K = \frac{N}{2}$ for typical sample selection in attribution distribution alignment.
For \modelname~model, we set the balance hyper-parameters as $\lambda_O = 0.5$ and $\lambda_A = 0.8$.
We set the hyper-parameters $\alpha = 0.1$ and $\nu = 0.1$ for solving the typical sample selection method and attribution subspace modelling respectively.
For all the experiments, we perform five random experiments and report the average results.
We choose Adam \cite{Adam} as optimizer, and adopt Hit Rate$@k$ (HR$@k$), Recall$@k$, and NDCG$@k$ \cite{kat} as the ranking evaluation metrics with $k = 10$.

\begin{figure*} 
    \centering
    
    \subfigure[\modelname-Base]{
    \begin{minipage}[t]{0.23\linewidth} 
    \includegraphics[width=4.5cm]{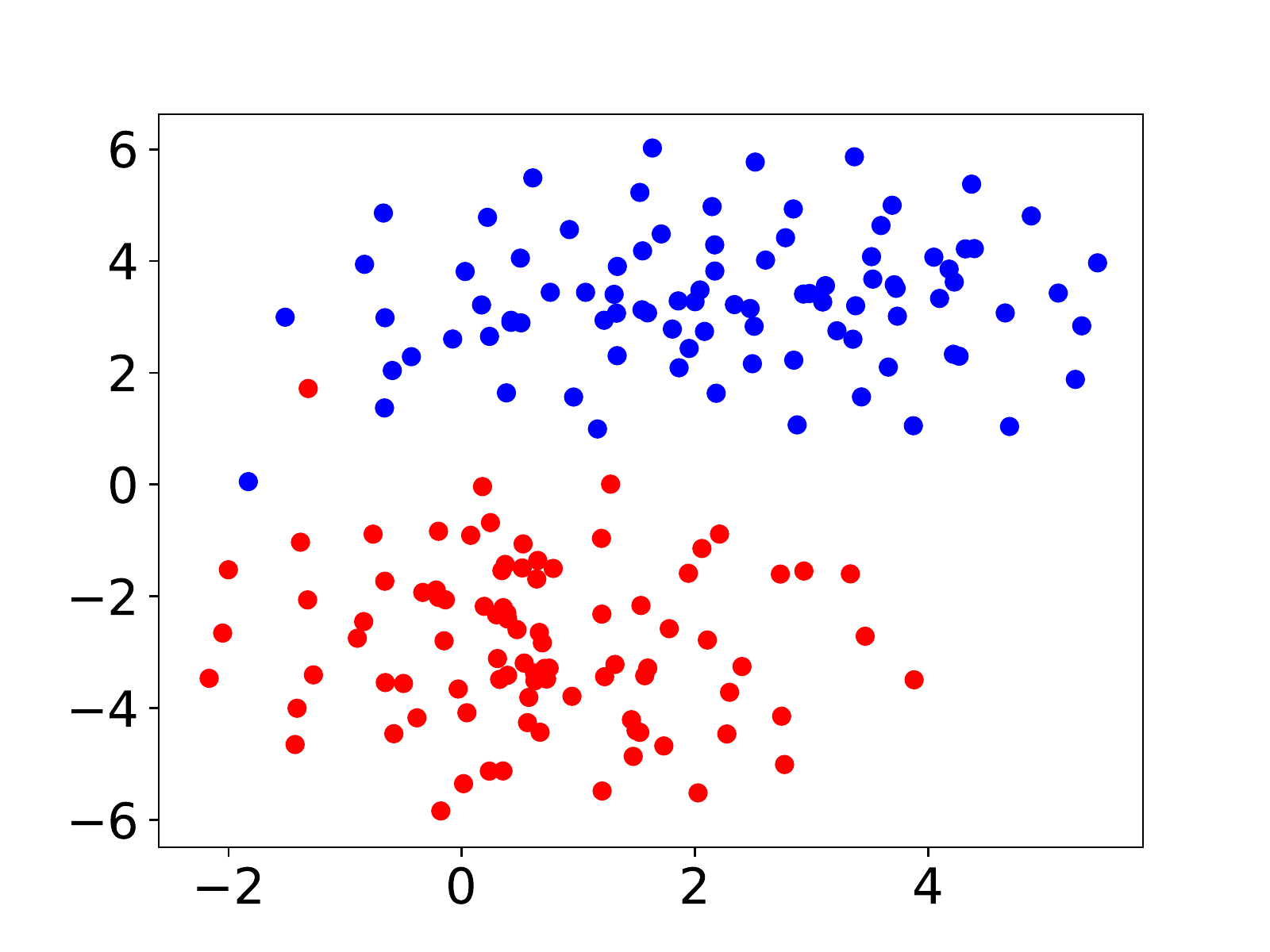}\\
    \includegraphics[width=4.5cm]{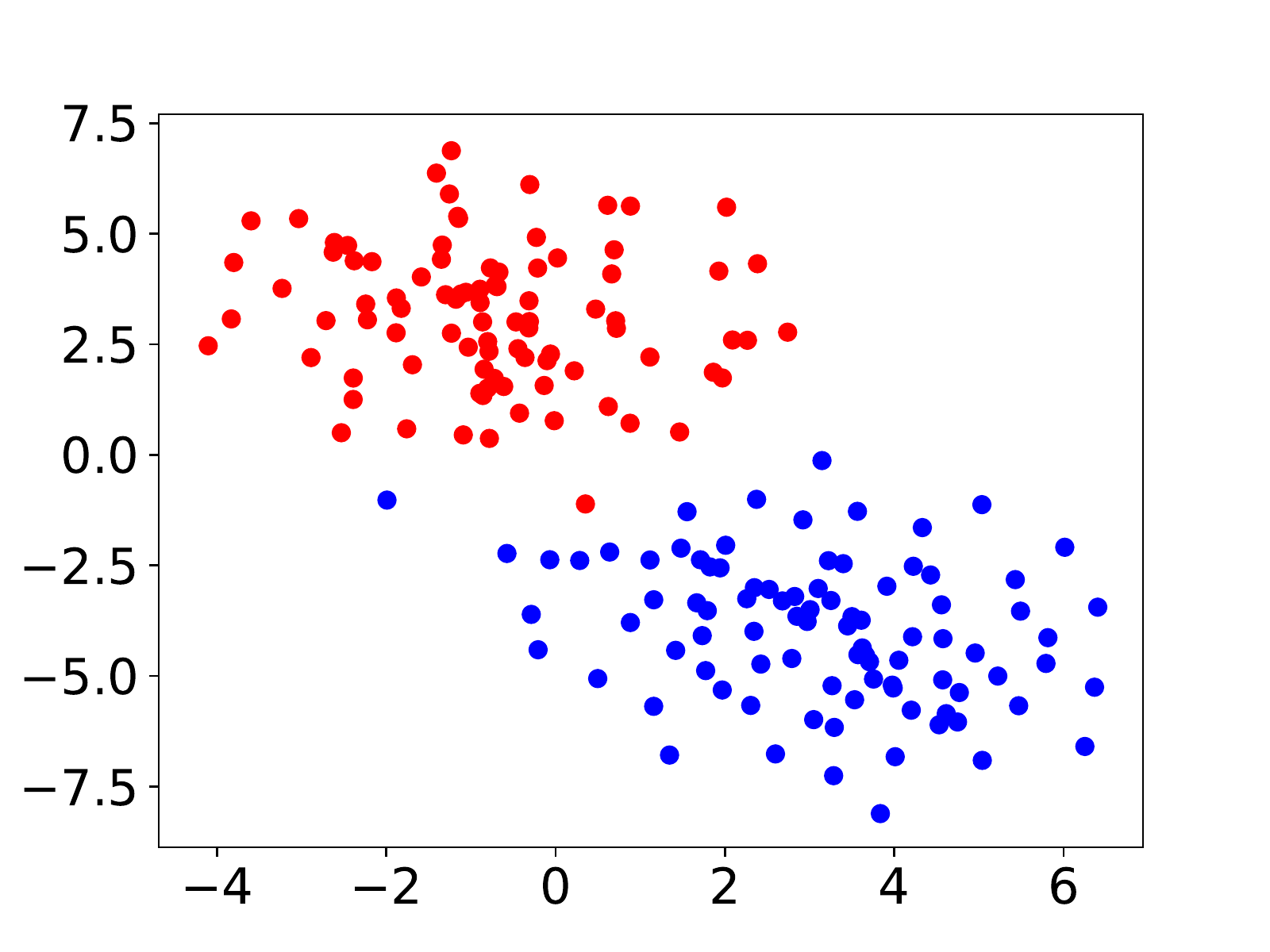}
    \end{minipage}
}
    \subfigure[DARec]{
    \begin{minipage}[t]{0.23\linewidth} 
    \includegraphics[width=4.5cm]{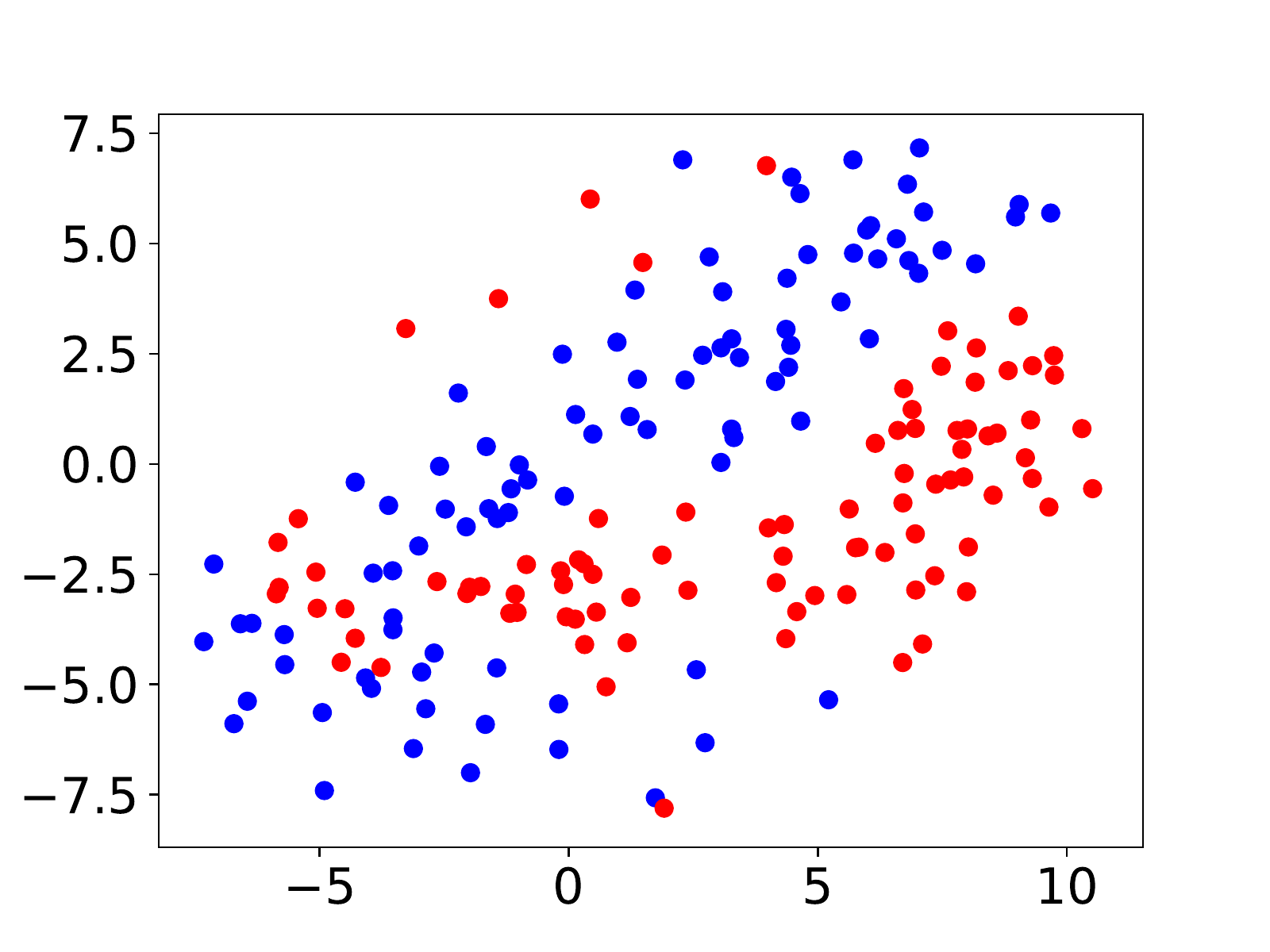}\\
    \includegraphics[width=4.5cm]{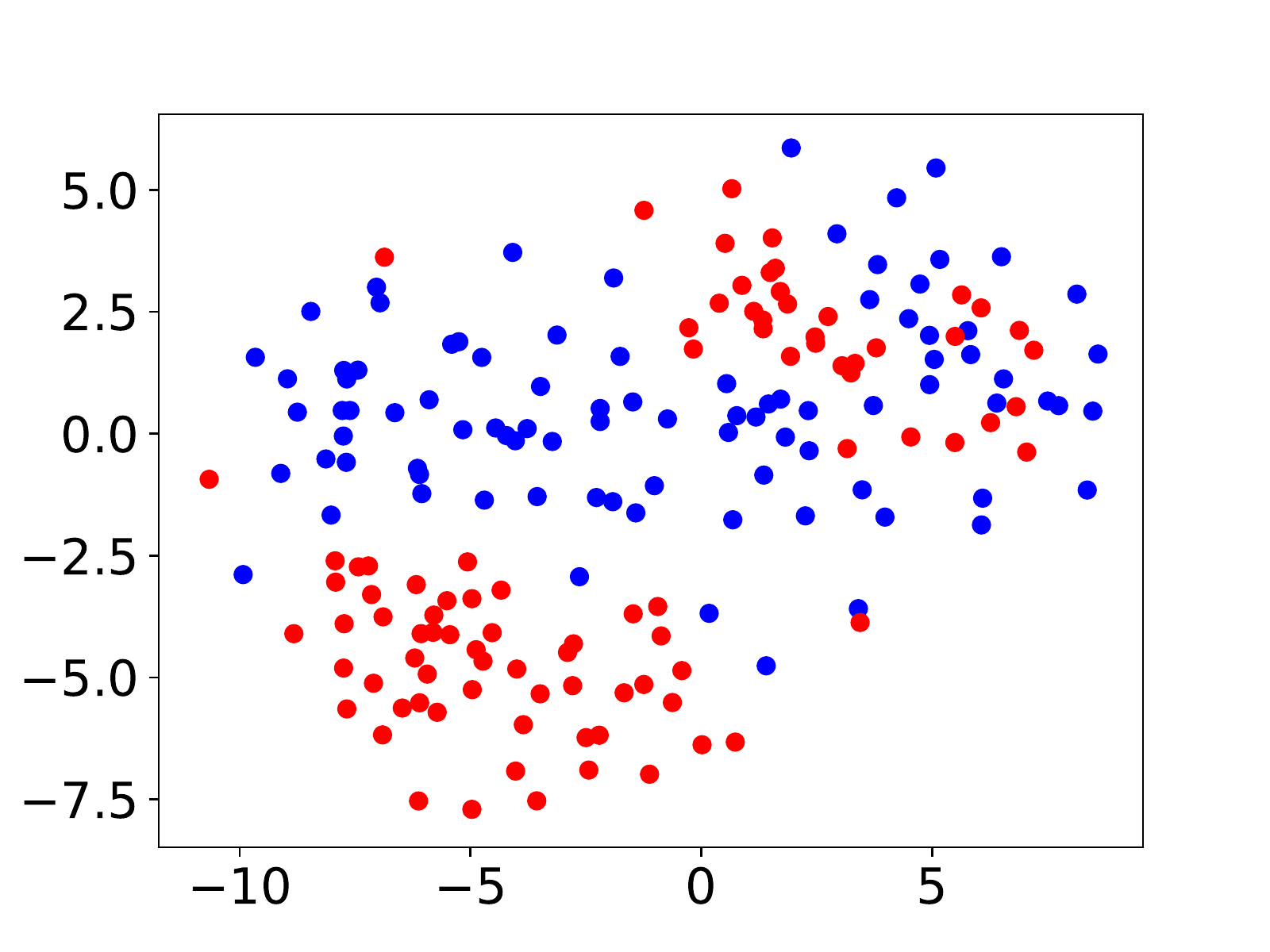}
    \end{minipage}
}
    \subfigure[ESAM]{
    \begin{minipage}[t]{0.23\linewidth} 
    \includegraphics[width=4.5cm]{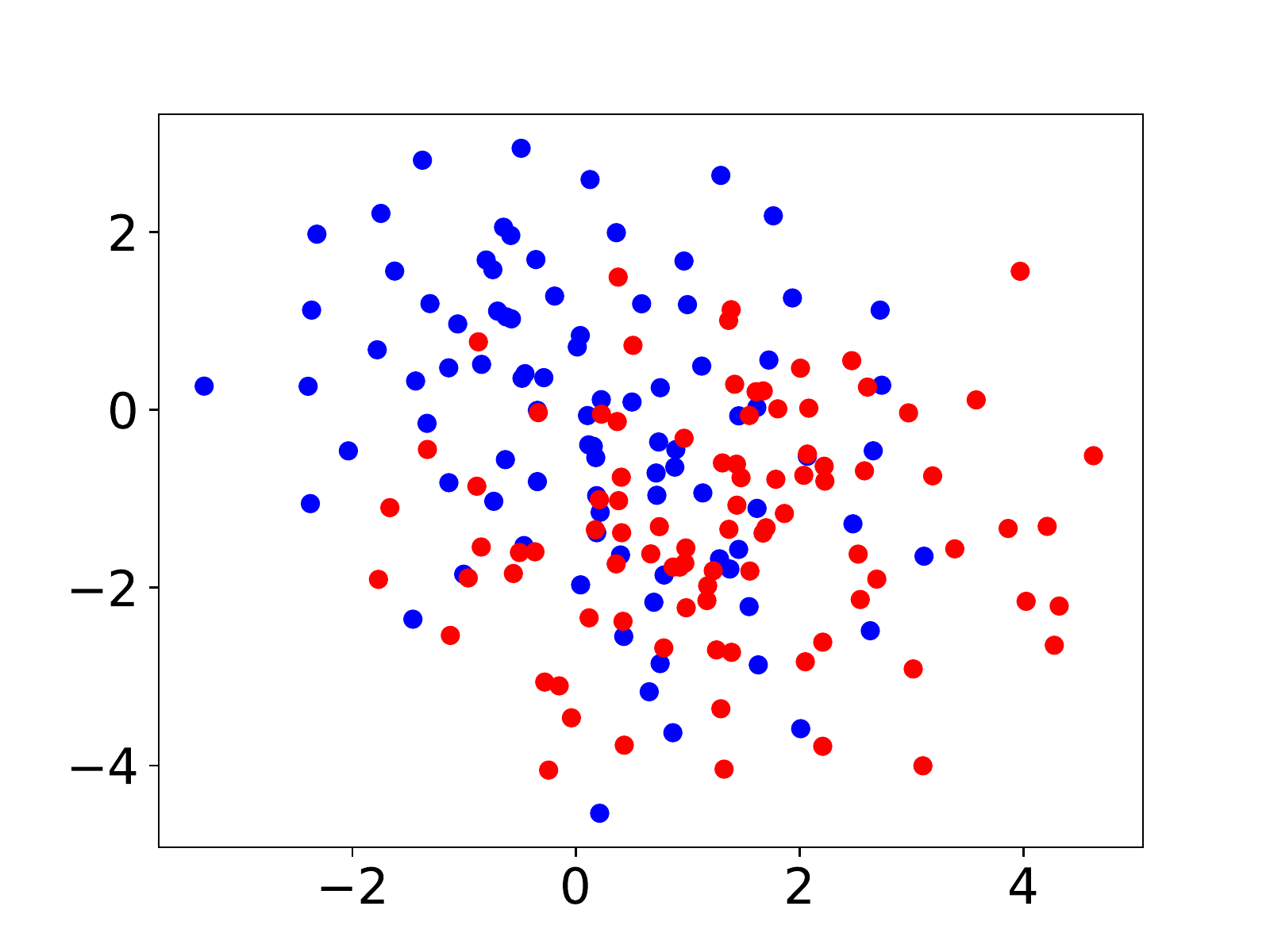}\\
    \includegraphics[width=4.5cm]{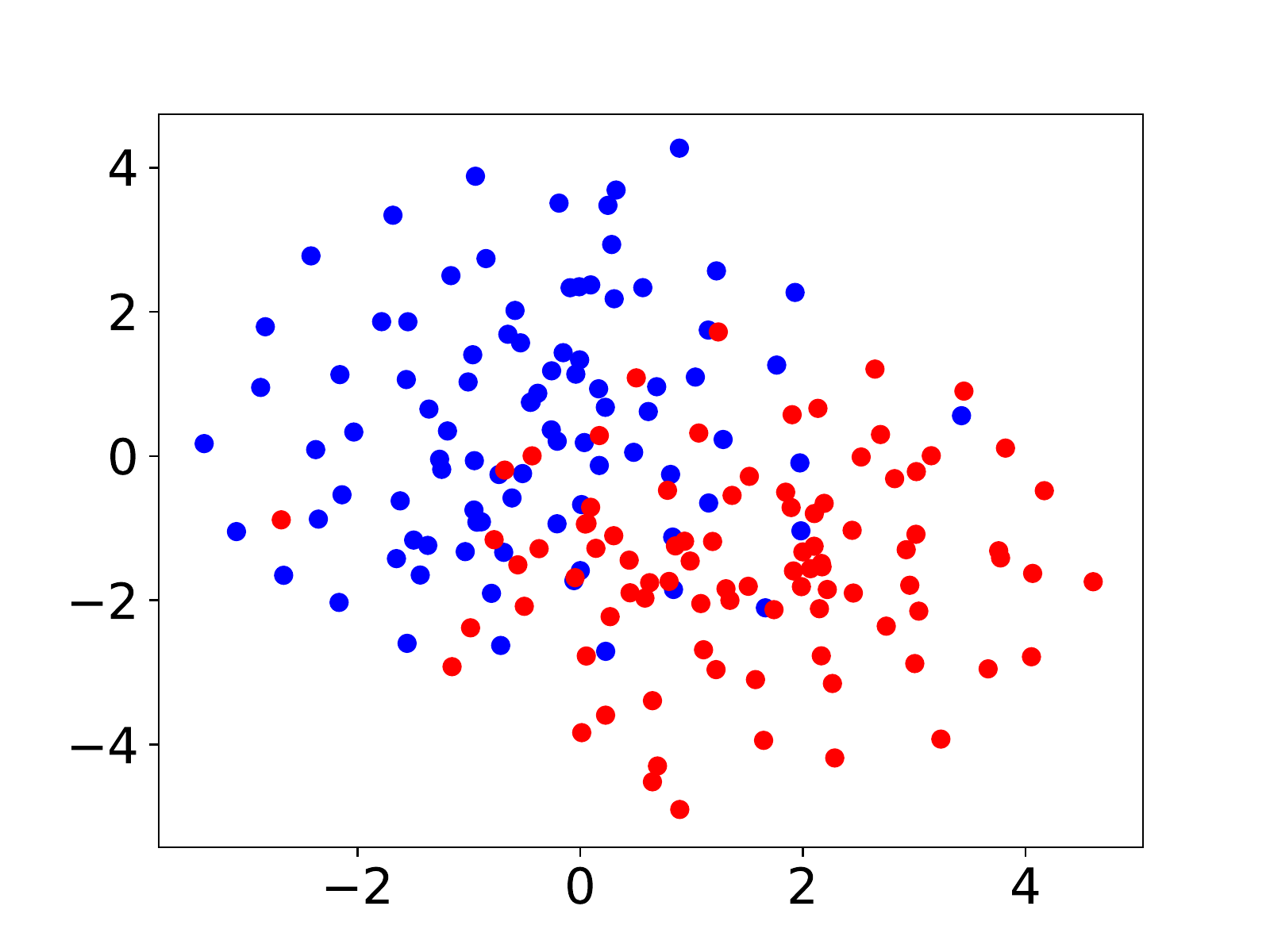}
    \end{minipage}
}
    \subfigure[\modelname]{
    \begin{minipage}[t]{0.23\linewidth} 
    \includegraphics[width=4.5cm]{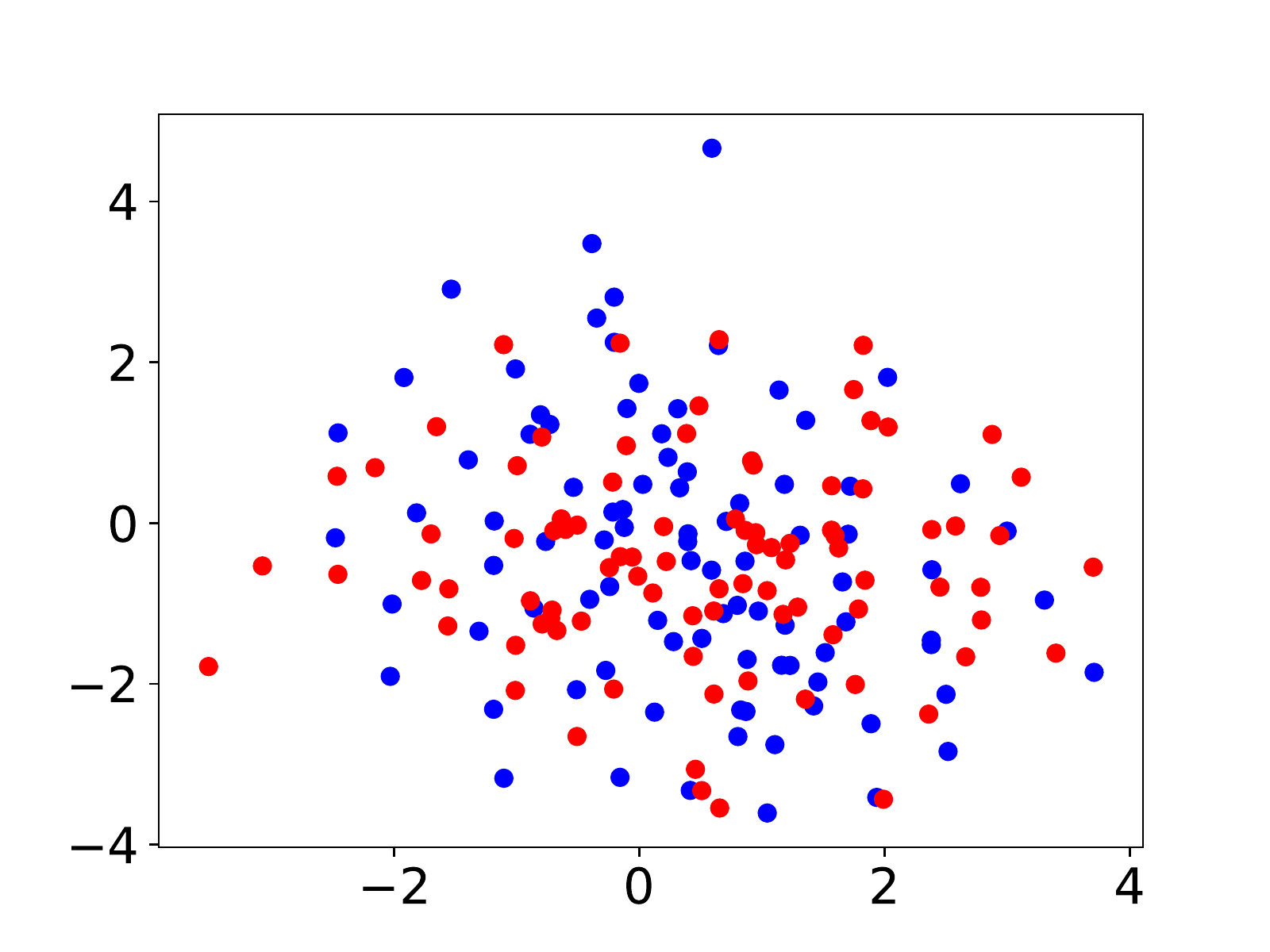}\\
    \includegraphics[width=4.5cm]{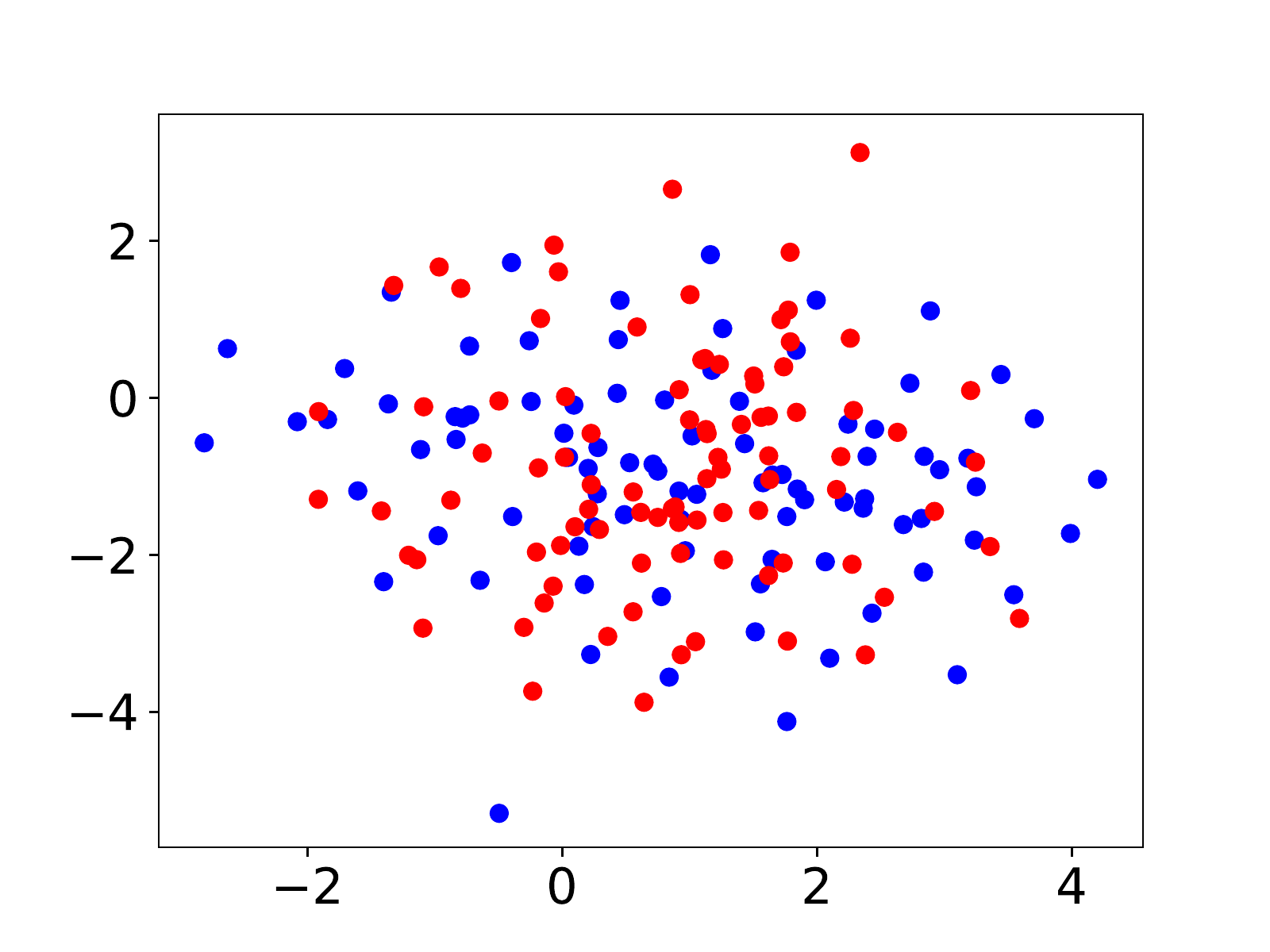}
    \end{minipage}
}

	  \caption{The t-SNE visualization of user latent embeddings on \textbf{Douban Movie} $\rightarrow$ \textbf{Douban Music} (first row) and item latent embeddings (second row). The user (item) latent embeddings in the source domain are shown with red dots and that in the target domain are shown with blue dots.}
	  \label{fig:tsne} 
\end{figure*}

\subsection{Baseline}
We compare our proposed \textbf{\modelname~}with the following state-of-the-art recommendation models.
(1) \textbf{DeepCoNN} \cite{deepconn} Deep Cooperative Neural Networks (DeepCoNN) is the first deep collaborative model to leverage both user and item textual features from reviews for recommendation.
(2) \textbf{NARRE} \cite{narre} Neural Attentional Rating Regression with Explanations (NARRE) utilizes two parallel CNNs with attention mechanism to extract review-level information for recommendation. 
(3) \textbf{ESCOFILT} \cite{esco} Extractive Summarization-Based Collaborative Filtering (ESCOFILT) is a BERT-based state-of-the-art collaborative filtering model based on user and item review information. 
(4) \textbf{Rec-DAN} \cite{recdan} Discriminative Adversarial Networks for Recommender System (Rec-DAN) adopts adversarial training strategy to align the joint user-item textual features to transfer useful knowledge.
(5) \textbf{TDAR} \cite{tdar} Text-enhanced Domain Adaptation (TDAR) is the state-of-the-art reviewed-based non-overlapped CDR model which adopts adversarial training strategy to user and item embeddings respectively with text memory network. 
(6) \textbf{ESAM} \cite{esam} Entire Space Adaptation Model (ESAM) adopts attribute correlation alignment to improve long-tail recommendation performance by suppressing inconsistent distribution between displayed and non-displayed items.
(7) \textbf{DARec} \cite{darec} Deep Domain Adaptation for Cross-Domain Recommendation via Transferring Rating Patterns (DARec) adopts adversarial training strategy to extract and transfer knowledge patterns for shared users across domains.
Note that the original \textbf{ESAM} and \textbf{DARec} models cannot be directly applied to RNCDR tasks, and thus we adopt the same rating prediction module as \textbf{\modelname}~for them.
Besides, for a fair comparison, all the models use the same types of data and pre-processing methods during experiments.


%

%

%





\begin{figure} 
    \centering
        \subfigure[Effects of $\lambda_{O}$]{
    \begin{minipage}[t]{0.47\linewidth} 
    \includegraphics[width=4.2cm]{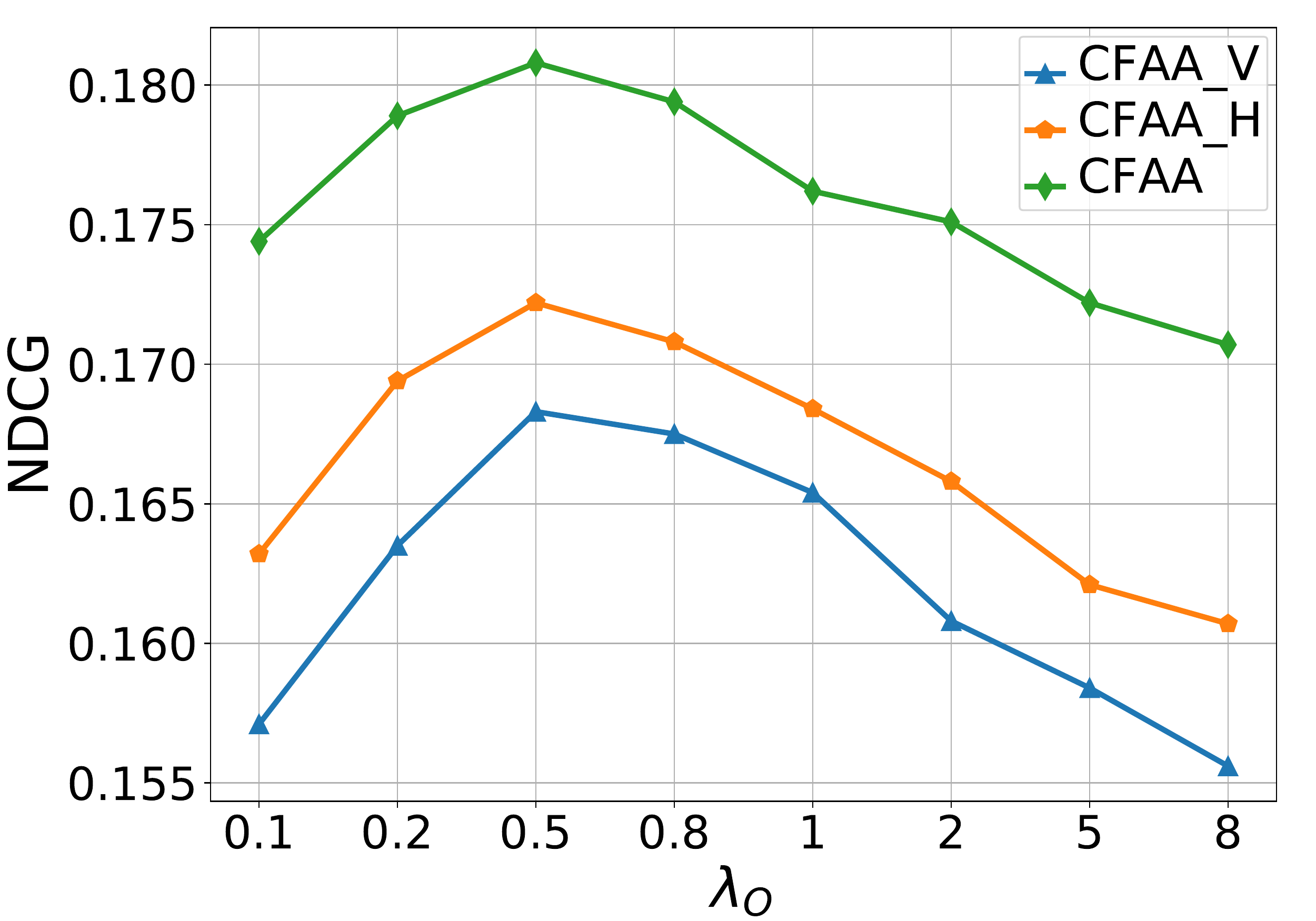}
    \end{minipage}
}
        \subfigure[Effects of $\lambda_{A}$]{
    \begin{minipage}[t]{0.47\linewidth} 
    \includegraphics[width=4.2cm]{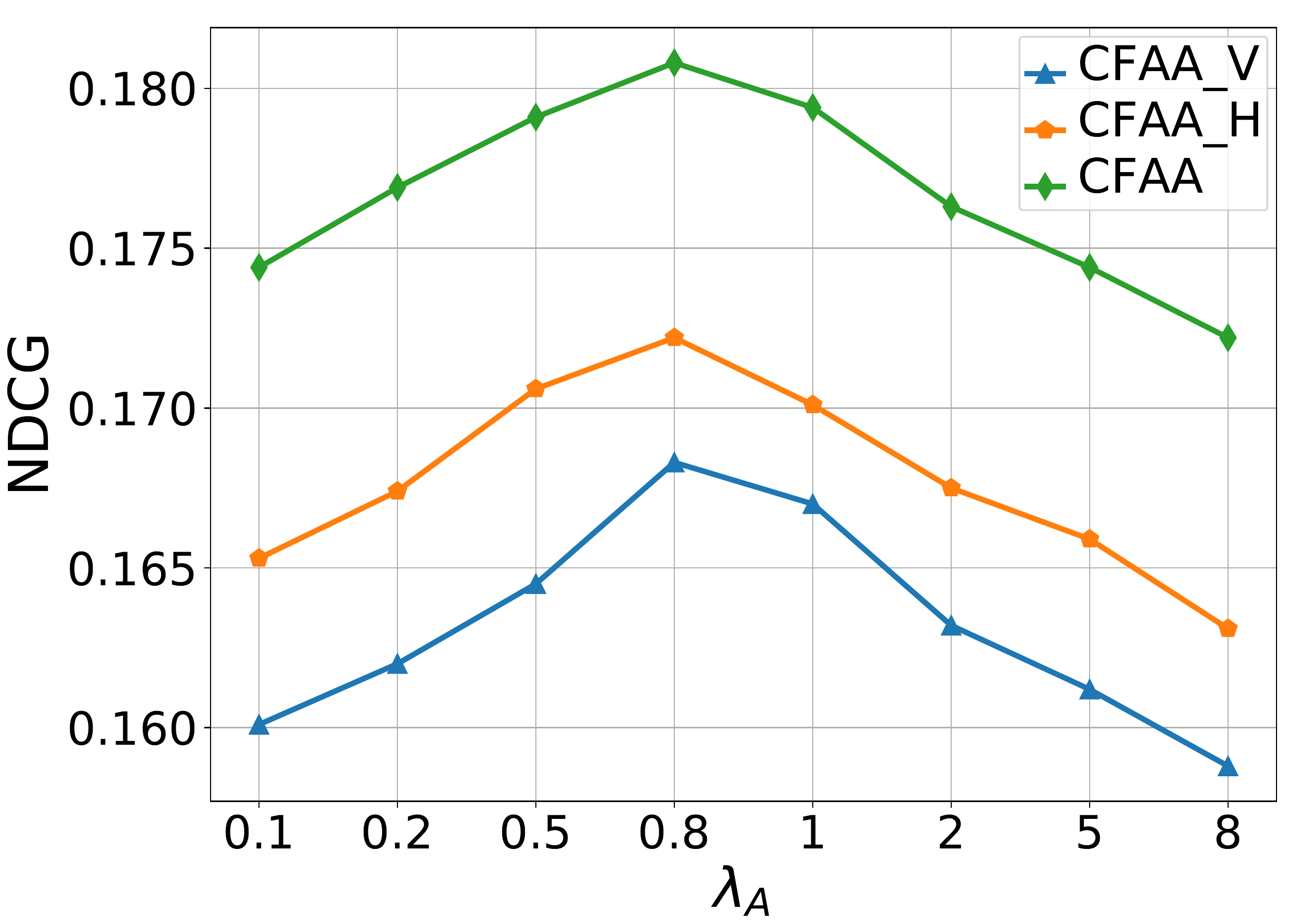}
    \end{minipage}
}
	  \caption{(a)-(b) show the effect of hyper-parameters $\lambda_{O}$ and $\lambda_{A}$ on model performance.}
	  \label{fig1:param}
\end{figure}

\subsection{Recommendation Performance (RQ1)}
\nosection{Results and discussion} 
The comparison results on Douban and Amazon datasets are shown in Table 1. 
Note that \modelname-Base represents the model that only adopts the rating prediction module for collaborative filtering without embedding attribution alignment.
From them, we can find that: (1) Only adopting the single target domain information (e.g., \textbf{DeepCoNN}) cannot obtain satisfying results under the RNCDR settings due to the data sparsity problem.
(2) \modelname-Base outperforms the previous reviewed-based recommendation model (e.g., \textbf{ESCOFILT}), indicating that adopting BERT-based review embedding with one-hot ID and multi-hot historical rating information can efficiently enhance the model performance.
(3) Although \textbf{ESAM} achieves better performance than \modelname-Base due to its attribution correlation congruence, it still fails to capture the nonlinear and complex topology structure among different attributions.
(4) Although \textbf{DARec} with gradient reverse layer can obtain good performance, the unstable adversarial training may hurdle the model to obtain more accurate results \cite{dirt-t}. 
As a result, it will eventually cause coarsely matching across domains. 
Meanwhile although \textbf{DARec} and \textbf{TDAR} both utilizes adversarial training, \textbf{DARec} equipped with more expressive user and item embeddings can obtain more delightful results. 
(5) \textbf{\modelname} consistently achieves the best performance, which proves that embedding attribution alignment module with attribution distribution and relevance alignment can significantly improve the prediction accuracy.
Notably that our proposed \textbf{\modelname} can enhance the performance when source and target domains are both similar (e.g., \textbf{Amazon Movie} $\rightarrow$ \textbf{Amazon Video}) and different (e.g., \textbf{Amazon Movie} $\rightarrow$ \textbf{Amazon Clothes}). 




\nosection{Visualization} 
To better show the user embeddings across domains, we visualize the t-SNE embeddings \cite{Laurens2008Visualizing} for \modelname-Base, \textbf{DARec}, \textbf{ESAM}, and \modelname.
The results of \textbf{Douban Movie} $\rightarrow$ \textbf{Douban Music} are shown in Fig.~\ref{fig:tsne}(a)-(d).
The first and second row denote user and item embeddings, respectively.
From it, we can see that (1) \modelname-Base cannot reduce the embedding bias and discrepancy on both users and items across domains, leading to insufficient knowledge transfer, as shown in Fig.~\ref{fig:tsne}(a).
(2) \textbf{ESAM} and \textbf{DARec} can marginally align user and item embedding attributions to a certain extent, but there still exists domain discrepancy which causes negative transfer, as shown in Fig.~\ref{fig:tsne}(b)-(c).
(3) \modelname~with attribution distribution and relevance alignment can better match users and items across domains, as shown in Fig.~\ref{fig:tsne}(d).
The visualization results illustrates the validity of our model.

\subsection{Analysis (RQ2 and RQ3)}

\nosection{Ablation}
To study how does each module of \textbf{\modelname~}contribute on the final performance, we compare \textbf{\modelname}~with its several variants, including \textbf{\modelname-V} and \textbf{\modelname-H}. 
\textbf{\modelname-V} only adopts the vertical attribution (i.e., attribution distribution) alignment while \textbf{\modelname-H} only adopts the horizontal attribution (i.e., attribution relevance) alignment.
The comparison results are shown in Table 1.
From it, we can observe that 
(1) \textbf{\modelname-V} and \textbf{\modelname-H} always get more accurate output predictions than \textbf{\modelname}-Base, which indicates that reducing embedding attribution alignment is essential. 
(2) However, \textbf{\modelname-V} and \textbf{\modelname-H} still cannot achieve the best results against \textbf{\modelname}.
Simply aligning vertical embedding probability distribution on \textbf{\modelname-V} will sometimes neglect the attribution relevance and cause negative transfer.
Likewise, only concentrating on the horizontal attribution alignment will sometimes ignore the attribution discrepancy across domains. 
Overall, the above ablation study demonstrates that our proposed embedding alignment module is effective in solving the RNCDR problem.

\begin{table}[!t]
\small
\centering
\caption{The results on $d_\mathcal{A}$ for domain discrepancy.}
\label{tab:douban1}
\begin{tabular}{ccccc}
\toprule
\multirow{2}{*}{}
& \multicolumn{2}{c}{(Amazon) Movie$\rightarrow$Music}
& \multicolumn{2}{c}{(Douban) Movie$\rightarrow$Music}\\
\cmidrule(lr){2-3} \cmidrule(lr){4-5}
& User &Item
&User &Item\\

\midrule
DARec & 1.5441 & 1.5294 & 1.4823 & 1.5045\\
ESAM & 1.5082 & 1.5310 & 1.4656 & 1.4780\\
\midrule 
\modelname-Base & 1.7503 & 1.7354 & 1.7372 & 1.7215\\
\modelname-V & 1.4125 & 1.4496 & 1.3839 & 1.3742\\
\modelname-H & 1.4039 & 1.4363 & 1.3950 & 1.3831\\
\modelname & \textbf{1.3548} & \textbf{1.3635} & \textbf{1.3084} & \textbf{1.2996} \\
\midrule

\end{tabular}
\vspace{-0.2cm} 
\end{table}

\nosection{Distribution Discrepancy}
The domain adaptation theory \cite{ben2007analysis} suggests proxy $\mathcal{A}$-distance as a measure of cross-domain discrepancy. 
We adopt $d_{\mathcal{A}}(\mathcal{S},$ $\mathcal{T})=2(1-2\epsilon(h))$ to analysis the distance between two domains, 
where $\epsilon(h)$ is the generalization error of a linear classifier $h$ that discriminates the source domain $\mathcal{S}$ and the target domain $\mathcal{T}$ \cite{ben2007analysis}. 
Table 2 demonstrates the domain discrepancy on \textbf{Amazon Movie} $\rightarrow$ \textbf{Amazon Music} and \textbf{Douban Movie} $\rightarrow$ \textbf{Douban Music} tasks using \modelname-Base, \textbf{\modelname-V}, \textbf{\modelname-H}, \modelname, and several baseline methods. 
From it, we can conclude that:
(1) The large number of $d_\mathcal{A}$ on \modelname-Base indicates the existence of embedding attribution discrepancy between the source and target domains. 
(2) Most of the current baselines (e.g.,\textbf{ESAM}) can reduce $d_\mathcal{A}$ but they still cannot obtain lower $d_\mathcal{A}$ than \modelname, indicating their limitations on domain adaptation.
(3) By adopting the embedding attribution alignment methods, \textbf{\modelname-V} and \textbf{\modelname-H} can reduce the embedding attribution discrepancy according to the $d_\mathcal{A}$ 
Furthermore, we can observe that \modelname~achieves the lowest $d_\mathcal{A}$ distance, because it considers both vertical and horizontal embedding attribution alignment. 
Lower $d_\mathcal{A}$ can always obtain better results since more useful knowledge can be transferred from source to target domains, which is consistent with the previous ablation study.

\nosection{Effect of hyper-parameters}
We finally study the effects of hyper-parameters on model performance.
For \modelname, we vary $\lambda_O$ and $\lambda_A$ in $\{0.1,0.2,0.5,0.8,1,2,5,8\}$ and report the results in Fig.~\ref{fig1:param}(a)-(b) on \textbf{Amazon Movie} $\rightarrow$ \textbf{Amazon Music}.
Fig.~\ref{fig1:param}(a)-(b) show the bell-shaped curves, indicating that choosing the proper hyper-parameters to balance the rating prediction loss and embedding attribution alignment loss can effectively improve the model performance.
Empirically, we choose $\lambda_O = 0.5$ and $\lambda_A = 0.8$.

\section{Conclusion}
In this paper, we propose Collaborative Filtering with Attribution Alignment model for solving review-based non-overlapped cross domain recommendation (\textbf{\modelname}), which includes the \textit{rating prediction module} and the \textit{embedding attribution alignment module}.
We innovatively adopt horizontal and vertical attribution alignment to better reduce the embedding discrepancy from different perspectives.
Vertical distribution alignment utilizes the typical sample selection with optimal transport to make them consistent across domains. 
Horizontal relevance alignment applies the subspace modelling with attribution graph alignment to reduce the discrepancy. 
We also conduct extensive experiments to demonstrate the superior performance of our proposed \modelname~on several datasets and tasks.
%


\begin{acks}
This work was supported in part by the National Key R\&D Program of China (No.72192823 and No.62172362).
\end{acks}

\bibliographystyle{ACM-Reference-Format}
\bibliography{main}

\clearpage

\appendix

\section{Typical Sample Selection Method}
As mentioned in Section 3.2.1, the typical sample selection algorithm is given by:
\begin{equation}
\myfont{
\begin{aligned}
\label{equ:multiple_proxies}
\min \ell_q = \sum_{i=1}^{N} \sum_{j=1}^{K}  (\boldsymbol{\Psi}^{Z^{\mathcal{X}}}_q )_{ij}\left(\boldsymbol{Z}_{iq}^\mathcal{X} - (\boldsymbol{M}^{\mathcal{X}}_Z)_{jq} \right)^2 + \alpha \cdot \boldsymbol{\mathcal{R}}\left((\boldsymbol{\Psi}^{Z^{\mathcal{X}}}_q)_{ij} \right)\\
s.t.\,\, (\boldsymbol{\Psi}^{Z^{\mathcal{X}}}_q )_{i} \boldsymbol{1} = 1, \boldsymbol{\mathcal{R}}\left((\boldsymbol{\Psi}^{Z^{\mathcal{X}}}_q)_{ij} \right) = (\boldsymbol{\Psi}^{Z^{\mathcal{X}}}_q)_{ij} \log (\boldsymbol{\Psi}^{Z^{\mathcal{X}}}_q)_{ij}, (\boldsymbol{\Psi}^{Z^{\mathcal{X}}}_q)_{ij} > 0.
\end{aligned}
}
\end{equation}
We now provide the optimization details on the typical-proxies algorithm.
Alternatively updating $\boldsymbol{M}^{\mathcal{X}}_Z$ and $\boldsymbol{\Psi}^{Z^{\mathcal{X}}}_{q}$ can solve Equation \eqref{equ:multiple_proxies} efficiently.

\nosection{Update $\boldsymbol{\Psi}^{Z^{\mathcal{X}}}_{q}$}
We first fix $\boldsymbol{M}^{\mathcal{X}}_Z$ and update $\boldsymbol{\Psi}^{Z^{\mathcal{X}}}_{q}$.
By using Lagrangian multiplier to minimize the objective function, we have:
\begin{equation}
\myfont{
\begin{aligned}
\label{equ:b_lagrang}
\ell_Q = \sum_{i=1}^{N} \sum_{j=1}^{K}  (\boldsymbol{\Psi}^{Z^{\mathcal{X}}}_q )_{ij}\zeta_{ijq} + \alpha \cdot \boldsymbol{\mathcal{R}}((\boldsymbol{\Psi}^{Z^{\mathcal{X}}}_q )_{ij}) + \chi \sum_{i=1}^{N} \left(\sum_{j=1}^{K} (\boldsymbol{\Psi}^{Z^{\mathcal{X}}}_q )_{ij} - 1 \right),
\end{aligned}
}
\end{equation}
where $\zeta_{ijq} = (\boldsymbol{Z}_{iq}^\mathcal{X} - (\boldsymbol{M}^{\mathcal{X}}_Z )_{jq} )^2$ and $\chi$ is the Lagrangian multiplier.
Taking the differentiation of Equation~\eqref{equ:b_lagrang} w.r.t. $(\boldsymbol{\Psi}^{Z^{\mathcal{X}}}_q )_{ij}$ and setting it to 0, we obtain:
\begin{equation}
\begin{aligned}
\label{equ:norm-grad}
\frac{\partial \ell_Q}{\partial (\boldsymbol{\Psi}^{Z^{\mathcal{X}}}_q )_{ij} } = \zeta_{ijq} + \alpha (\log (\boldsymbol{\Psi}^{Z^{\mathcal{X}}}_q )_{ij} + 1) + \chi = 0.
\end{aligned}
\end{equation}
By solving and simplifying Equation \eqref{equ:norm-grad}, we have:
\begin{equation}
\begin{aligned}
\label{equ:step1}
(\boldsymbol{\Psi}^{Z^{\mathcal{X}}}_q )_{ij} = \exp \left( -\frac{\alpha + \chi }{\alpha} \right) \exp \left( -\frac{\zeta_{ijq}}{\alpha} \right).
\end{aligned}
\end{equation}
Meanwhile, taking $\sum\limits_{j=1}^K (\boldsymbol{\Psi}^{Z^{\mathcal{X}}}_q )_{ij} = 1$ into Equation \eqref{equ:step1}, we have:
\begin{equation}
\begin{aligned}
\label{equ:norm}
\sum\limits_{j=1}^K \exp \left( -\frac{\alpha + \chi }{\alpha} \right) \exp \left( -\frac{\zeta_{ijq}}{\alpha} \right) = \exp \left( -\frac{\alpha + \chi }{\alpha} \right) \sum\limits_{j=1}^K \exp \left( -\frac{\zeta_{ijq}}{\alpha} \right)=  1.
\end{aligned}
\end{equation}
That is,
\begin{equation}
\begin{aligned}
\exp \left( -\frac{\alpha + \chi }{\alpha} \right) = \frac{1}{\sum\limits_{j=1}^K \exp \left( -\frac{\zeta_{ijq}}{\alpha} \right)}.
\end{aligned}
\end{equation}
Thus, the final solution of $(\boldsymbol{\Psi}^{Z^{\mathcal{X}}}_q )_{ij}$ is given by:
\begin{equation}
\begin{aligned}
\label{equ:find_s2}
(\boldsymbol{\Psi}^{Z^{\mathcal{X}}}_q)_{ij} = \frac{\exp \left( -\zeta_{ijq}/ \alpha \right)}{\sum_{k=1}^K \exp ( -\zeta_{ikq}/ \alpha )}.
\end{aligned}
\end{equation}
\nosection{Update $\boldsymbol{M}^{\mathcal{X}}_Z$}
After we have updated $\boldsymbol{\Psi}^{Z^{\mathcal{X}}}_{q}$, we fix it as a constant and update $\boldsymbol{M}^{\mathcal{X}}_Z$. Thus, Equation \eqref{equ:b_lagrang} becomes
\begin{equation}
\begin{aligned}
\label{equ:m}
\min_{\boldsymbol{M}^{\mathcal{X}}_Z}  \sum_{i=1}^{N} \sum_{j=1}^{K}  (\boldsymbol{\Psi}^{Z^{\mathcal{X}}}_q)_{ij}||\boldsymbol{Z}_{iq}^\mathcal{X} - (\boldsymbol{M}^{\mathcal{X}}_Z)_{jq}||_2^2. 
\end{aligned}
\end{equation}
Taking the differentiation of Equation \eqref{equ:m} w.r.t. $\boldsymbol{M}^{\mathcal{X}}_Z$ and setting it to 0, we can update $\boldsymbol{M}^{\mathcal{X}}_Z$ as:
\begin{equation}
\begin{aligned}
\label{equ:find_s3}
(\boldsymbol{M}^{\mathcal{X}}_Z)_{jq} = \frac{\sum_{i=1}^{N} (\boldsymbol{\Psi}^{Z^{\mathcal{X}}}_q )_{ij}\boldsymbol{Z}_{iq}^\mathcal{X} }{\sum_{i=1}^{N} (\boldsymbol{\Psi}^{Z^{\mathcal{X}}}_q)_{ij}}.
\end{aligned}
\end{equation}
We can obtain the stable solution of $\boldsymbol{\Psi}^{Z^{\mathcal{X}}}_{q}$ and $\boldsymbol{M}^{\mathcal{X}}_Z$ through several iterations.

\begin{table}[!t]
\small
\centering
\caption{Statistics on Douban and Amazon datasets.}

\begin{tabular}{ccccc}
\toprule
Datasets  & Items & Users  & Interactions  & Density\\
\midrule
Amazon Movie (S) & 50,052 & 123,960 & 1,697,532 & 0.027\%\\
Amazon Book (S) & 43,168 & 95,643 & 1,032,019 & 0.025\%\\
Douban Movie (S) & 34,893 & 151,258 & 1,278,401 & 0.024\% \\
Douban Book (S) & 38,776 & 111,270 & 965,041 & 0.022\% \\
\midrule

Amazon Music (T) & 7,710 & 11,053 & 106,188 & 0.124\%\\
Amazon Video (T) & 1,580 & 4,555 & 11,137 & 0.155\%\\
Amazon Clothes (T) & 21,554 & 35,669 & 89,176 & 0.012\%\\
Douban Music (T) & 3,562 & 11,278 & 2,1451 & 0.053\% \\

\bottomrule
\end{tabular}

\label{table:dataset}
\end{table} 

\section{Attribution Subspace Modelling}
As mentioned in Section 3.2.2, the attribution subspace modelling algorithm is given by:
\begin{equation}
\begin{aligned}
\label{equ:slim_2}
\min_{\boldsymbol{B}^\mathcal{X}_Z,\boldsymbol{\Phi}} \frac{1}{2}\left| \left|\boldsymbol{Z}^\mathcal{X} - \boldsymbol{Z}^\mathcal{X}\boldsymbol{B}^\mathcal{X}_Z \right|\right|_2^2 + \nu {\rm Tr}\left( (\boldsymbol{B}^\mathcal{X}_Z)^T \boldsymbol{\Phi} \boldsymbol{B}^\mathcal{X}_Z  \right)\\
s.t.\,\, {\rm diag}\left(\boldsymbol{B}^\mathcal{X}_Z\right) = 0, \boldsymbol{\Phi} = \left( \boldsymbol{B}^\mathcal{X}_Z (\boldsymbol{B}^\mathcal{X}_Z)^T \right)^{-\frac{1}{2}}.
\end{aligned}
\end{equation}
Alternatively updating $\boldsymbol{B}^\mathcal{X}_Z$ and $\boldsymbol{\Phi}$ can solve Equation \eqref{equ:slim_2} efficiently.

We first fix $\boldsymbol{\Phi}$ and update $\boldsymbol{B}^\mathcal{X}_Z$.
By using Lagrangian multiplier to minimize the objective function, we have:
\begin{equation}
\begin{aligned}
\label{equ:lagrang}
\ell_B = \frac{1}{2}\left| \left|\boldsymbol{Z}^\mathcal{X} - \boldsymbol{Z}^\mathcal{X}\boldsymbol{B}^\mathcal{X}_Z \right|\right|_2^2 + \nu {\rm Tr}\left( (\boldsymbol{B}^\mathcal{X}_Z)^T \boldsymbol{\Phi} \boldsymbol{B}^\mathcal{X}_Z  \right) + \boldsymbol{\gamma} {\rm diag}(\boldsymbol{B}^\mathcal{X}_Z).
\end{aligned}
\end{equation}
Taking the differentiation of Equation \eqref{equ:lagrang} w.r.t. $\boldsymbol{B}^\mathcal{X}_Z$ and setting it to 0, we obtain:
\begin{equation}
\begin{aligned}
\label{equ:B-grad}
\frac{\partial \ell_B}{\partial \boldsymbol{B}^\mathcal{X}_Z} = ( \boldsymbol{Z}^\mathcal{X})^T (  \boldsymbol{Z}^\mathcal{X}\boldsymbol{B}^\mathcal{X}_Z - \boldsymbol{Z}^\mathcal{X} ) + \nu \boldsymbol{\Xi}\boldsymbol{B}^\mathcal{X}_Z  +{\rm diagMat}(\boldsymbol{\gamma}) = 0,
\end{aligned}
\end{equation}
where $\boldsymbol{\Phi} + \boldsymbol{\Phi}^T = \boldsymbol{\Xi}$.
${\rm diagMat}(\cdot)$ denotes the diagonal matrix.
By solving and simplifying Equation \eqref{equ:B-grad}, we have:
\begin{equation}
\begin{aligned}
\boldsymbol{B}^\mathcal{X}_Z = ( ( \boldsymbol{Z}^\mathcal{X})^T \boldsymbol{Z}^\mathcal{X} +  \nu \boldsymbol{\Xi} )^{-1} ( ( \boldsymbol{Z}^\mathcal{X})^T \boldsymbol{Z}^\mathcal{X} -{\rm diagMat}(\boldsymbol{\gamma}) ).
\end{aligned}
\end{equation}
Here, we suppose that $\nu$ is always sufficient large and $( ( \boldsymbol{Z}^\mathcal{X})^T \boldsymbol{Z}^\mathcal{X} +  \nu \boldsymbol{\Xi} )^{-1}$ is invertible \cite{emb}. 
We define $\boldsymbol{\Theta} = ( ( \boldsymbol{Z}^\mathcal{X})^T \boldsymbol{Z}^\mathcal{X} +  \nu \boldsymbol{\Xi} )^{-1}$ and substitute it into Equation \eqref{equ:B-grad}:
\begin{equation}
\begin{aligned}
\boldsymbol{B}^\mathcal{X}_Z = \boldsymbol{I} - \boldsymbol{\Theta} \cdot {\rm diagMat}(\nu \boldsymbol{1} + \boldsymbol{\gamma}).
\end{aligned}
\end{equation}
The unknown value of $\boldsymbol{\gamma}$ can be solved by the diagonal constraint ${\rm diag}(\boldsymbol{B}^\mathcal{X}_Z ) = 0$.
We can obtain that $\boldsymbol{\gamma} = \boldsymbol{1} \oslash {\rm diag}(\boldsymbol{\Xi})$, where $\oslash$ denotes the elementwise division.
Thus, the final solution of $\boldsymbol{B}^\mathcal{X}_Z$ is given by:
\begin{equation}
\begin{aligned}
\left(\boldsymbol{B}^\mathcal{X}_Z\right)_{ij} = 
\left\{
\begin{aligned}
&0, \quad \quad \quad \quad i = j\\
& -\frac{\boldsymbol{\Theta}_{ij}}{\boldsymbol{\Theta}_{jj}},\quad \rm{Others}.
\end{aligned}
\right.
\end{aligned}
\end{equation}
After we have updated $\boldsymbol{B}^\mathcal{X}_Z$, we fix it as a constant and update $\boldsymbol{\Phi}$ through the equality constraint $\boldsymbol{\Phi} = ( \boldsymbol{B}^\mathcal{X}_Z (\boldsymbol{B}^\mathcal{X}_Z)^T)^{-\frac{1}{2}}$.
Utilizing this iteration method until it converges, we can obtain the results of $\boldsymbol{B}^\mathcal{X}_Z$ and $\boldsymbol{\Phi}$.

\section{Datasets and Tasks} 
We conduct extensive experiments on two popularly used real-world datasets, i.e., \textit{Douban} and \textit{Amazon}.
The \textbf{Douban} dataset includes Book, Music, and Movie and the \textbf{Amazon} dataset has five domains, i.e., Movies and TV (Movie), Books (Book), CDs and Vinyl (Music), Instant Videos (Video) and Clothes (Clothes).
%
The detailed statistics of these datasets after pre-process are shown in Table 3.

\end{document}